\documentclass[11pt,a4paper]{article}
\usepackage{amsmath}
\usepackage{amsfonts}
\usepackage{diagbox}
\usepackage{array}
\usepackage{bm}
\usepackage{tikz}
\usetikzlibrary{fit}					
\usetikzlibrary{backgrounds}	%

\usepackage{mathrsfs}
\usepackage{cite}

\newcommand*\xbar[1]{%
  \hbox{%
    \vbox{%
      \hrule height 0.5pt 
      \kern0.5ex
      \hbox{%
        \kern-0.1em
        \ensuremath{#1}%
        \kern-0.1em
      }%
    }%
  }%
}

\begin{document}

\title{Structure of renormalization constants for theories with multiple couplings in the MS-like subtraction schemes}

\author{
G.V.Kovyrshin\,${}^{a}$, N.P.Meshcheriakov\,${}^{b}$, V.V.Shatalova\,${}^{c}$, and K.V.Stepanyantz${}^{d}$ $\vphantom{\Big(}$
\medskip\\
{\small{\em Moscow State University,}}\\
\vphantom{1}\vspace*{-2mm}\\
${}^a${\small{\em Faculty of Physics, Department of Quantum Statistics and Field Theory, 119991,}}\\
${}^b${\small{\em Faculty of Physics, Department of Quantum Theory and High Energy Physics, 119991,}}\\
${}^c${\small{\em AESC MSU – Kolmogorov boarding school, Department of Physics, 119192,}}\\
${}^d${\small{\em Faculty of Physics, Department of Theoretical Physics, 119991,}}\\
\vphantom{1}\vspace*{-2mm}\\
{\small{\em Moscow, Russia.}}\\
\vphantom{1}\vspace*{-2mm}}

\maketitle

\begin{abstract}
For theories with multiple couplings we construct simple expressions for the four-dimensional (or, in general, integer-dimensional) renormalization constants assuming that all divergences are logarithmical. These expressions allow relating all coefficients at $\varepsilon$-poles, logarithms, and (if exist) mixed terms to the coefficients of the renormalization group functions in any order of the perturbation theory for MS-like renormalization prescriptions. The result admits such a formulation in that $\varepsilon$-poles and $\ln\Lambda/\mu$ enter on the same footing. For theories with two and three couplings we present explicit expressions for the pole/logarithm structure of renormalization constants in the lowest orders of the perturbation theory. They are verified by comparisons with the two-loop explicit calculation for ${\cal N}=1$ SQCD+SQED and also with the previously known three-loop calculations for the $\varphi^4$-theory with two couplings.
\end{abstract}

\allowdisplaybreaks

\section{Introduction}
\hspace*{\parindent}

It is well known that quantum field theory models should be regularized because quantum corrections are in general divergent in the ultraviolet region. For renormalizable theories these divergences can be removed by renormalizing various couplings and fields, see, e.g., \cite{Bogolyubov:1980nc,Collins:1984xc}. Certainly, the most popular method of regularization is dimensional regularization \cite{'tHooft:1972fi,Bollini:1972ui,Ashmore:1972uj,Cicuta:1972jf}. In this case divergent contributions to quantum corrections turn into poles in $\varepsilon\equiv 4-D$ (or $\varepsilon=D_0-D$, where $D_0$ is an integer space-time dimension). However, sometimes dimensional regularization is not convenient, and other regularization \cite{Jack:1997sr,Gnendiger:2017pys} may be useful. For instance, dimensional regularization explicitly breaks supersymmetry \cite{Delbourgo:1974az}. Therefore, in the supersymmetric case, its modification, called dimensional reduction \cite{Siegel:1979wq}, is usually used. Nevertheless, some features of supersymmetric theories are not seen with this regularization, and Slavnov's higher covariant derivative regularization \cite{Slavnov:1971aw,Slavnov:1972sq,Slavnov:1977zf} (in the superfield formulation \cite{Krivoshchekov:1978xg,West:1985jx}) is more preferable in this case, see \cite{Stepanyantz:2019lyo} for detail. For instance, it is this regularization that allows to construct the perturbative all-loop derivation of the Novikov, Shifman, Vainshtein, and Zakharov (NSVZ) exact $\beta$-function \cite{Novikov:1983uc,Jones:1983ip,Novikov:1985rd,Shifman:1986zi} and a simple prescription giving an NSVZ scheme in all orders \cite{Stepanyantz:2016gtk,Stepanyantz:2019ihw,Stepanyantz:2020uke}, while in the $\overline{\mbox{DR}}$ scheme the NSVZ equation is not valid \cite{Jack:1996vg,Jack:1996cn,Jack:1998uj}. The higher covariant derivative regularization belongs to the regularizations of the cutoff type. The regularizations of this type are formulated in integer dimension and contain a dimensionful regularization parameter $\Lambda$ which effectively cuts the loop integration in the ultraviolet region. In this case divergences appear as powers of $\ln\Lambda/\mu$, where $\mu$ is a renormalization point. It has long been noted that these logarithms are analogous to the $\varepsilon$-poles for the dimensional regularization (reduction). For instance, the coefficients at simple $\varepsilon$-poles and at the first power of $\ln\Lambda/\mu$ differ in the numbers of loops \cite{Chetyrkin:1980sa}. The coefficients at higher $\varepsilon$-poles can be expressed in terms of the coefficients at simple $\varepsilon$-poles with the help of 't Hooft pole equations \cite{tHooft:1973mfk}.\footnote{Analogous equations can be written even for the nonrenormalizable theories, see, e.g., \cite{Kazakov:1987jp,Kazakov:2016wrp,Borlakov:2016mwp,Kazakov:2019wce,Solodukhin:2020vuw}.} Similarly, for the cutoff type regularizations the coefficients at higher powers of $\ln\Lambda/\mu$ are related to the coefficients at the first powers of this logarithm, see, e.g., \cite{Derkachev:2017nhd,Meshcheriakov:2022tyi}. These relations are very important for making multiloop calculations, because analyzing the results for the coefficients at higher $\varepsilon$-poles or higher logarithms, it is possible to verify the correctness. Therefore, taking into account that the multiloop calculations are usually very complicated, it is desirable to understand the dependence of the renormalization constants on both higher $\varepsilon$-poles and higher logarithms. Note that the dependence of the renormalization constants on higher powers of $\varepsilon$-poles and $\ln\Lambda/\mu$ is quite different \cite{Meshcheriakov:2023fmk}, so that it seems rather interesting to establish the correspondence between these two regularization parameters. For this purpose, it is convenient to use such a regularization which generates both $\varepsilon$-poles and logarithms. It can be constructed by a special modification of the standard dimensional technique. Namely, if the $D$-dimensional bare coupling $\widetilde\alpha_0$ is presented as $\widetilde\alpha_0 = \alpha_0\Lambda^\varepsilon$, where $\alpha_0$ is dimensionless and the parameter $\Lambda$ has the dimension of mass, then it is possible to take $\Lambda\ne \mu$ and $\Lambda\to\infty$. In this case the dependence of the renormalization constants on $\ln\Lambda/\mu$ will be exactly the same as for the regularization of the cutoff type (certainly if one omit $\varepsilon$-poles and the mixed terms). That is why this regularization can be considered as a bridge connecting two different types of regularizations. Some explicit multiloop calculations involving this technique have been done in \cite{Aleshin:2015qqc,Aleshin:2016rrr,Aleshin:2019yqj}. For theories with a single coupling the coefficients at higher $\varepsilon$-poles, higher logarithms and the mixed terms can be related to the coefficients at simple poles and the first power of $\ln\Lambda/\mu$ (which are in turn related to the coefficients in the renormalization group functions (RGFs)) by the equations derived in \cite{Meshcheriakov:2023fmk}. Subsequently, these equations have been rewritten in a simple and beautiful form in \cite{Meshcheriakov:2024qwj}, see also \cite{Stepanyantz:2025akv}. In this paper we construct the analogous representation of the renormalization constants for theories with multiple couplings. The resulting equations encode all relations between poles and logarithms for regularizations of any type supplemented with minimal subtraction of divergences (possibly, modified by rescaling of $\Lambda$ or $\mu$). Moreover, they can be presented in such a form where $\varepsilon$-poles and logarithms enter on the same footing. This establishes the correspondence between $\varepsilon$-poles and logarithms in all orders for any renormalizable theory under the assumption that the divergences are logarithmical.

The paper is organized as follows. In Sect.~\ref{Section_Regularization} we introduce a version of the dimensional technique for which divergences contain both $\varepsilon$-poles and logarithms. It will allow obtaining the results valid for the regularizations of both dimensional and cutoff types. Next, in Sect.~\ref{Section_Exact_Expressions} for theories with multiple couplings we derive the (exact in all-orders) expressions for the renormalization constants which relate them to RGFs and generate all coefficients at higher $\varepsilon$-poles, logarithms, and mixed terms. In particular, we will see that it is possible to write these expressions in such a form that $\varepsilon^{-1}$ and $\ln\Lambda/\mu$ enter the corresponding equations in a very similar way. Some explicit formulas for theories with two and three couplings in the lowest loops (following from the general expressions) are presented in Sect.~\ref{Section_Explicit_Result} and Appendix~\ref{Appendix_Explicit_Z}. The technique and general equations considered in this paper are illustrated by an explicit calculation in Sect.~\ref{Section_Gamma}. Namely, in this section (using the regularization by dimensional reduction with $\Lambda\ne \mu$) we calculate the two-loop renormalization constant for the matter superfields in ${\cal N}=1$ SQCD+SQED in the $\overline{\mbox{DR}}$ scheme and demonstrate that the resulting expressions are in agreement with the general expressions obtained in the previous sections. With the help of these general expressions, in Sect.~\ref{Section_Kleinert}, we also verify the results of the three-loop calculations for the $\varphi^4$-theory with two couplings presented in \cite{Kleinert:2001ax} and demonstrate that the general equations really reproduce the structure of the higher $\varepsilon$-poles in the explicit three-loop expressions for the renormalization constants of this theory. A brief summary and discussion is given in Conclusion. Some auxiliary calculations and lengthy expressions are presented in Appendices.

\section{Regularization and definitions of RGFs}
\hspace*{\parindent}\label{Section_Regularization}

Let us consider a theory that contains $n$ dimensionless (in the integer dimension $D_0$) bare couplings denoted by $\alpha_{i0}$, where the index $i$ numerates couplings, and the subscript $0$ indicates that these couplings are bare. For regularizing a theory with the help of dimensional regularization, it is formally considered in the space of the non-integer dimension $D$. Then the divergences of loop integrals in the dimension $D_0$ (e.g., $D_0=4$) will turn into $\varepsilon$-poles, where

\begin{equation}
\varepsilon \equiv D_0 - D,
\end{equation}

\noindent
e.g., $\varepsilon\equiv 4-D$ for the ordinary four-dimensional theories. Moreover, the couplings that are dimensionless in $D_0$ acquire in this case the dimension $m^{\varepsilon}$. However, it is more convenient to deal with the dimensionless couplings. To construct them, we need to introduce a dimensionful regularization parameter $\Lambda$, and present the original dimensionful couplings $\widetilde \alpha_{i0}$ in the form

\begin{equation}
\widetilde\alpha_{i0}\equiv \Lambda^\varepsilon \alpha_{i0},
\end{equation}

\noindent
where the couplings $\alpha_{i0}$ are dimensionless. The regularization parameter $\Lambda$ is arbitrary. Standardly, it is set equal to the renormalization scale $\mu$. However, in general it is not necessary. For instance, we may define the renormalized $D$-dimensional couplings $\bm{\alpha}_i$ according to the prescription

\begin{equation}\label{AlphaD_Renormalization}
\alpha_{i0} = \Big(\frac{\mu}{\Lambda}\Big)^{\varepsilon} \bm{\alpha}_i\, \bm{Z}_{\alpha_i}^{-1}(\bm{\alpha}, \varepsilon^{-1})
\end{equation}

\noindent
and assume that $\Lambda\to \infty$. The renormalized couplings $\bm{\alpha}_i$ certainly depend on the renormalization scale $\mu$. However, any physical quantity (e.g., decay probabilities or cross-sections) should be independent of the renormalization scale $\mu$. Therefore, the scale dependence of the coefficients in the perturbative series must compensate the scale dependence of renormalized couplings, so that the physical observables depend only on the integral curve of the renormalization group equations. However, if the perturbative series is truncated at a certain order, the scheme dependence will appear in the following orders, see \cite{Ioffe:2010zz} and references therein. In this case a good choice of the renormalization scale is desirable. For instance, in scattering problems it is convenient to choose the parameter $\mu$ of the same order as typical momentum transfer of the process.

The presence of the ratio $\Lambda/\mu$ in Eq. (\ref{AlphaD_Renormalization}) is convenient because (as we will see in what follows) this will allow constructing the dependence of the renormalization constants not only on higher poles but also on higher logarithms. Taking into account that the equations relating the coefficients at (pure) higher logarithms to the coefficients of RGFs are regularization independent (although the coefficients of RGFs can certainly be different), the result for pure logarithms will be valid for any regularization of the cutoff type.

The renormalization should also be made for masses and fields. The corresponding $D$-dimensional renormalization constants $\bm{Z}$ are determined by requiring the finiteness of the corresponding renormalized Green functions in the limit $\varepsilon\to 0$. For instance, the two-point function of a certain field\footnote{For simplicity, all couplings on which it depends we will denote by $\alpha$.}

\begin{equation}\label{GD_Renormalization}
G_R\Big(\bm{\alpha},\ln\frac{\mu}{P}\Big) = \lim\limits_{\varepsilon\to 0}\bm{Z}(\bm{\alpha},\varepsilon^{-1})\, G\Big[\Big(\frac{\mu}{P}\Big)^\varepsilon\bm{\alpha} \bm{Z}_\alpha^{-1}(\bm{\alpha},\varepsilon^{-1}),\varepsilon^{-1}\Big],
\end{equation}

\noindent
where $P$ is the (absolute value of the Euclidean) momentum, should be finite in this limit. It is convenient to encode divergences in the RGFs. The $D$-dimensional RGFs are defined by the equations

\begin{eqnarray}\label{RGFsD_Definition}
&& \bm{\beta}_i(\bm{\alpha},\varepsilon) \equiv \frac{d\bm{\alpha}_i(\alpha_0(\Lambda/\mu)^\varepsilon,\varepsilon^{-1})}{d\ln\mu}\bigg|_{\alpha_0=\text{const}};\qquad\nonumber\\
&& \bm{\gamma}(\bm{\alpha},\varepsilon) \equiv \frac{d\ln \bm{Z}(\bm{\alpha}, \varepsilon^{-1})}{d\ln\mu} \bigg|_{\alpha_0=\text{const}} = \sum\limits_{i=1}^n \bm{\beta}_i(\bm{\alpha},\varepsilon) \frac{\partial \ln \bm{Z}}{\partial \bm{\alpha}_i}.\qquad
\end{eqnarray}

In the minimal subtraction (MS) renormalization scheme, by definition, the renormalization constants contain only $\varepsilon$-poles, $\bm{Z}(\alpha,\varepsilon^{-1}\to 0)\to 1$. There are also some MS-like renormalization prescriptions, e.g., the $\overline{\mbox{MS}}$ scheme \cite{Bardeen:1978yd}, which in our conventions is obtained for $\Lambda = \mu\,\exp(\gamma/2)/\sqrt{4\pi}$, where $\gamma\equiv - \Gamma'(1)$ is the Euler constant. It is should also be mentioned that for the MS-like subtraction schemes the renormalization constants do not depend on masses \cite{tHooft:1973mfk,Collins:1973yy,Joglekar:1986be,Joglekar:1989ev}.

Note that, alternatively, the renormalization can be made in a different (integer-dimensional) way. In this case the renormalization of couplings and Green functions is made according to the prescription

\begin{eqnarray}\label{Alpha4_Renormalization}
&& \frac{1}{\alpha_{i0}} = \frac{Z_{\alpha_i} (\alpha, \varepsilon^{-1}, \ln \Lambda/\mu)}{\alpha_i};\\
\label{G_4_Renormalization}
&& G_R\Big(\alpha,\ln\frac{\mu}{P}\Big) = \lim\limits_{\varepsilon\to 0} Z(\alpha,\varepsilon^{-1},\ln\Lambda/\mu)\, G\Big[\Big(\frac{\Lambda}{P}\Big)^\varepsilon\alpha Z_\alpha^{-1}(\alpha,\varepsilon^{-1},\ln\Lambda/\mu), \varepsilon^{-1}\Big],\qquad
\end{eqnarray}

\noindent
where the renormalization constants $Z_{\alpha_i}$ and $Z$ include $\varepsilon$-poles, powers of $\ln\Lambda/\mu$, and mixed terms of the structure $\varepsilon^{-p}\ln^q\Lambda/\mu$, where $p$ and $q$ are positive integers. In this paper we will consider the MS-like renormalization prescriptions for that the renormalization constants do not contain finite terms,

\begin{equation}
Z(\alpha,\varepsilon^{-1}\to 0,\ln\frac{\Lambda}{\mu}\to 0) \to 1.
\end{equation}

Note that the four- (or, in general, integer-) dimensional renormalization constants do not include positive powers of $\varepsilon$, which are very essential in the $D$-dimensional renormalization. Therefore, (for the same bare couplings $\alpha_{i0}$) the $D$-dimensional renormalized couplings $\bm{\alpha}_i$ differ from the integer-dimensional renormalized couplings $\alpha_i$. Due to this difference, we denote the former couplings in bold.

The four-/integer-dimensional RGFs are defined by the equations

\begin{equation}\label{RGFs4_Definition}
\beta_i(\alpha) \equiv \frac{d \alpha_i(\alpha_0, \varepsilon^{-1}, \ln \Lambda/\mu) }{d \ln \mu}\bigg|_{\alpha_0=\text{const}};\qquad
\gamma(\alpha) \equiv \frac{d\ln Z(\alpha,\varepsilon^{-1},\ln\Lambda/\mu)}{d\ln\mu}\bigg|_{\alpha_0=\text{const}}.
\end{equation}

\noindent
For the MS-like renormalization prescriptions these RGFs are related to the $D$-dimensional RGFs defined in Eq. (\ref{RGFsD_Definition}) by the equations \cite{Narison:1980ti} (see also \cite{Spiridonov:1984br,Kazakov:2008tr})

\begin{equation}\label{RGFs_Relations}
\bm{\beta}_i(\alpha,\varepsilon) = \beta_i(\alpha) - \varepsilon\alpha_i;\qquad \bm{\gamma}(\alpha,\varepsilon) = \gamma(\alpha).
\end{equation}

\noindent
For completeness, we present the derivation of these equations for theories with multiple couplings in Appendix \ref{Appendix_RGFs_Relations}. Eq. (\ref{RGFs_Relations}) implies that for the MS-like renormalization prescriptions the functions $\bm{\gamma}(\alpha,\varepsilon)$ and $\bm{\beta}_i(\alpha,\varepsilon)+\varepsilon\alpha_i$ do not in fact depend on $\varepsilon$. Note that for arbitrary (non-MS-like) subtraction schemes the equations (\ref{RGFs_Relations}) are not in general valid.

According to Eqs. (\ref{AlphaD_Renormalization}) and (\ref{Alpha4_Renormalization}), the anomalous dimensions for the renormalization of couplings are related to the corresponding $\beta$-functions by the equations

\begin{eqnarray}\label{Gamma_D_RG_Equation}
&& \bm{\gamma}_{\alpha_i}(\bm{\alpha},\varepsilon) = \frac{d\ln \bm{Z}_{\alpha_i}(\bm{\alpha}, \varepsilon^{-1})}{d\ln\mu} \bigg|_{\alpha_0=\text{const}} = \varepsilon + \frac{\bm{\beta}_i(\bm{\alpha},\varepsilon)}{\bm{\alpha}_i};\\
\label{Gamma_4_RG_Equation}
&& \gamma_{\alpha_i}(\alpha) = \frac{d\ln Z_{\alpha_i}(\alpha, \varepsilon^{-1},\ln\Lambda/\mu)}{d\ln\mu} \bigg|_{\alpha_0=\text{const}} = \frac{\beta_i(\alpha)}{\alpha_i}.
\end{eqnarray}

\noindent
This implies that it is not necessary to consider separately the renormalization of couplings because the corresponding renormalization constants can be obtained from general equations by substituting the anomalous dimensions (\ref{Gamma_D_RG_Equation}) or (\ref{Gamma_4_RG_Equation}).

\section{All-loop expressions for the renormalization constants}
\hspace*{\parindent}\label{Section_Exact_Expressions}

Let us construct explicit expressions for the renormalization constants relating them directly to RGFs. We will start with obtaining the dependence on $\ln\Lambda/\mu$. For this purpose, we rewrite the renormalization group equation in the form

\begin{eqnarray}\label{Ln_Z_RG_Equation}
&& \gamma(\alpha) = \frac{d}{d\ln\mu} \ln Z\Big(\alpha,\varepsilon^{-1},\ln\frac{\Lambda}{\mu}\Big)\bigg|_{\alpha_0=\text{const}}\nonumber\\
&&\qquad\qquad\qquad\qquad = \sum\limits_{i=1}^n \beta_i(\alpha) \frac{\partial}{\partial\alpha_i} \ln Z\Big(\alpha,\varepsilon^{-1},\ln\frac{\Lambda}{\mu}\Big) + \frac{\partial}{\partial\ln\mu}\ln Z\Big(\alpha,\varepsilon^{-1},\ln\frac{\Lambda}{\mu}\Big),\qquad
\end{eqnarray}

\noindent
where the partial derivative with respect to $\ln\mu$ in the right hand side is calculated at fixed values of all renormalized couplings $\alpha_i$. Then, the solution of the resulting equation

\begin{equation}
\frac{\partial}{\partial\ln\mu} Z = \Big(\gamma(\alpha) - \sum\limits_{i=1}^n \beta_i(\alpha)\frac{\partial}{\partial\alpha_i}\Big)Z
\end{equation}

\noindent
with the boundary condition $Z(\alpha,\varepsilon^{-1},0) = \bm{Z}(\alpha,\varepsilon^{-1})$, where $\bm{Z}(\alpha,\varepsilon^{-1})$ is the renormalization constant for the standard version of the dimensional regularization/reduction, can be presented in the form

\begin{equation}\label{Z_Ln_Dependence}
Z\Big(\alpha,\varepsilon^{-1},\ln\frac{\Lambda}{\mu}\Big) = \exp\bigg\{\ln\frac{\Lambda}{\mu}\Big( \sum\limits_{i=1}^n \beta_i(\alpha)\frac{\partial}{\partial\alpha_i} - \gamma(\alpha)\Big)\bigg\}\, \bm{Z}(\alpha,\varepsilon^{-1}).
\end{equation}

\noindent
The explicit expression for $\bm{Z}(\alpha,\varepsilon^{-1})$ will be constructed in what follows. However, before doing this, let us note that the same derivation can be repeated for any regularization of the cutoff type supplemented by the minimal subtraction of logarithms (MSL) renormalization prescription which corresponds to the boundary condition $Z(\alpha,\ln\Lambda/\mu\to 0)\to 1$ \cite{Kataev:2013eta}.\footnote{If the MSL renormalization prescription supplements the Higher covariant Derivative (HD) regularization, then we obtain the HD+MSL scheme \cite{Shakhmanov:2017wji,Stepanyantz:2017sqg}, which was numerously used for various multiloop calculations in supersymmetric theories (including the ones with multiple couplings), see, e.g., \cite{Kazantsev:2018nbl,Kuzmichev:2019ywn,Aleshin:2020gec,Aleshin:2022zln,Haneychuk:2022qvu,Shirokov:2022jyd,Shirokov:2023jya,Haneychuk:2025ehb}.}. In this case there are no $\varepsilon$-poles, so that the final expression for the renormalization constant $Z$ can be written in the form

\begin{equation}\label{Z_CutOff_MSL_Scheme}
Z\Big(\alpha,\ln\frac{\Lambda}{\mu}\Big) = \exp\bigg\{\ln\frac{\Lambda}{\mu}\Big( \sum\limits_{i=1}^n \beta_i(\alpha)\frac{\partial}{\partial\alpha_i} - \gamma(\alpha)\Big)\bigg\}\cdot 1,
\end{equation}

\noindent
where the operator in the argument of the exponential function is equal to the generator of rescaling transformations multiplied by $-\ln\Lambda/\mu$ \cite{Kataev:2024xbl}.

Now, let us proceed to obtaining an equation analogous to (\ref{Z_CutOff_MSL_Scheme}) for the regularizations of the dimensional type. For this purpose, we start with the second renormalization group equation in (\ref{RGFsD_Definition}) written in terms of the four-dimensional RGFs (with the help of Eq. (\ref{RGFs_Relations})),

\begin{equation}\label{Gamma_Equation_DR}
\gamma(\alpha) =  \sum\limits_{i=1}^n \Big(\beta_i(\alpha)-\varepsilon\alpha_i\Big) \frac{\partial \ln \bm{Z}(\alpha,\varepsilon^{-1})}{\partial \alpha_i}.
\end{equation}

\noindent
To find a solution of this equation, we introduce an auxiliary parameter $t$ by making the substitution $\alpha_i\to t\alpha_i$ simultaneously for all couplings. Taking into account that

\begin{equation}
\frac{\partial}{\partial \ln t}\ln \bm{Z}(t\alpha,\varepsilon^{-1}) = \sum\limits_{i=1}^n \alpha_i \frac{\partial\ln \bm{Z}(t\alpha,\varepsilon^{-1})}{\partial\alpha_i},
\end{equation}

\noindent
it is possible to present Eq. (\ref{Gamma_Equation_DR}) in the equivalent form

\begin{equation}\label{RG_Equation_DR}
\frac{\partial}{\partial t} \bm{Z}(t\alpha,\varepsilon^{-1}) = \frac{1}{t\varepsilon} \Big(\frac{1}{t}\sum\limits_{i=1}^n\beta_i(t\alpha)\frac{\partial}{\partial\alpha_i} - \gamma(t\alpha) \Big) \bm{Z}(t\alpha,\varepsilon^{-1}).
\end{equation}

\noindent
For $t=0$ we in fact obtain the free theory for that all couplings vanish, and, therefore, the renormalization constants are equal to 1. The value $t=1$ corresponds to the original theory. In this case it is also necessary to fix a renormalization prescription by fixing a finite part of the renormalization constant. Here we use MS scheme, for which only the $\varepsilon$-poles are included into the renormalization constant $\bm{Z}$. Therefore, Eq. (\ref{RG_Equation_DR}) should be supplemented with the boundary conditions

\begin{equation}\label{Boundary_Conditions_DR}
\bm{Z}(0,\varepsilon^{-1}) = 1;\qquad \bm{Z}(\alpha,\varepsilon^{-1}\to 0)\to 1,
\end{equation}

\noindent
where the {\it formal} limit $\varepsilon^{-1}\to 0$ corresponds to omitting all $\varepsilon$-poles. The solution of Eq. (\ref{RG_Equation_DR}) that satisfies the conditions (\ref{Boundary_Conditions_DR}) can be presented in the form

\begin{eqnarray}\label{Z_DR_Result}
\bm{Z}(\alpha,\varepsilon^{-1}) = T\exp\bigg\{\int\limits_0^1 \frac{dt}{t\varepsilon}\Big( \frac{1}{t}\sum\limits_{i=1}^n \beta_i(t\alpha)\frac{\partial}{\partial\alpha_i} - \gamma(t\alpha)\Big)\bigg\}\cdot 1,
\end{eqnarray}

\noindent
where $T$ denotes the ordering operator,

\begin{equation}
T\Big(A(t_1) A(t_2)\Big) = \left\{
\begin{array}{l}
{\displaystyle A(t_1) A(t_2),\qquad \mbox{if}\quad t_1>t_2;}\\
\vphantom{1}\\
{\displaystyle A(t_2) A(t_1),\qquad \mbox{if}\quad t_2>t_1.}
\end{array}
\right.
\end{equation}

Combining Eq. (\ref{Z_Ln_Dependence}) and (\ref{Z_DR_Result}) we obtain the general result for a renormalization constant in the case of using the version of dimensional regularization described in Sect. \ref{Section_Regularization} supplemented by an MS-like renormalization prescription,

\begin{eqnarray}\label{Z_Result_General}
&& Z\Big(\alpha,\varepsilon^{-1},\ln\frac{\Lambda}{\mu}\Big) = \exp\bigg\{\ln\frac{\Lambda}{\mu}\Big( \sum\limits_{i=1}^n \beta_i(\alpha)\frac{\partial}{\partial\alpha_i} - \gamma(\alpha)\Big)\bigg\}\,\nonumber\\
&&\qquad\qquad\qquad\qquad\qquad\qquad\qquad\quad
\times\, T\exp\bigg\{\smash{\int\limits_0^1} \frac{dt}{t\varepsilon}\Big( \frac{1}{t}\sum\limits_{i=1}^n \beta_i(t\alpha)\frac{\partial}{\partial\alpha_i} - \gamma(t\alpha)\Big)\bigg\}\cdot 1.\qquad
\end{eqnarray}

Certainly, the ordered exponential in the expressions (\ref{Z_DR_Result}) and (\ref{Z_Result_General}) is defined in the standard way as

\begin{eqnarray}\label{T_Exponent}
&& T\exp\bigg\{\int\limits_0^1 dt A(t)\bigg\} = \sum\limits_{k=0}^\infty \frac{1}{k!} \int\limits_0^1 dt_1 \int\limits_0^1 dt_2 \ldots \int\limits_0^1 dt_k T\Big(A(t_1) A(t_2)\ldots A(t_k)\Big)\nonumber\\
&&\qquad\qquad\qquad\qquad\qquad\qquad\qquad = \sum\limits_{k=0}^\infty \int\limits_0^1 dt_1 \int\limits_0^{t_1} dt_2 \ldots \int\limits_0^{t_{k-1}} dt_k A(t_1) A(t_2)\ldots A(t_k).\qquad
\end{eqnarray}

\noindent
It is also possible to present this expression in an equivalent form introducing the operator $\int\limits^\wedge dt$ that, by definition, acts on everything on the right of it according to the equation

\begin{equation}\label{Integral_Operator_Definition}
\int\limits^\wedge dt\, t^k \equiv \frac{t^{k+1}}{(k+1)}, \qquad k\ge 0,
\end{equation}

\noindent
and the product of integrals with hats implies that the upper limit of integration in the next integral is the variable of integration in the previous one. Then, taking into account that

\begin{equation}
\frac{1}{1-x} = \sum\limits_{k=0}^\infty x^k,
\end{equation}

\noindent
it is possible to present the ordered exponential in the form

\begin{equation}
T\exp\bigg\{\int\limits_0^1 dt A(t)\bigg\} = \Big(1-\int\limits^\wedge dt\, A(t) \Big)^{-1}\bigg|_{t=1}.
\end{equation}

\noindent
With the help of this equality the expression (\ref{Z_Result_General}) can be rewritten in the form

\begin{eqnarray}\label{Z_Result_Calculable}
&& Z\Big(\alpha,\varepsilon^{-1},\ln\frac{\Lambda}{\mu}\Big) = \exp\bigg\{\ln\frac{\Lambda}{\mu}\Big( \sum\limits_{i=1}^n \beta_i(\alpha)\frac{\partial}{\partial\alpha_i} - \gamma(\alpha)\Big)\bigg\}
\nonumber\\
&&\qquad\qquad\qquad\qquad\qquad\quad
\times\bigg(1- \int\limits^\wedge \frac{dt}{t\varepsilon}\Big( \frac{1}{t}\sum\limits_{i=1}^n \beta_i(t\alpha)\frac{\partial}{\partial\alpha_i} - \gamma(t\alpha)\Big)\bigg)^{-1}\cdot 1\,\Bigg|_{t=1},\qquad
\end{eqnarray}

\noindent
which is the most convenient for constructing the perturbative expansion. All explicit expressions presented below in Sect. \ref{Section_Explicit_Result} and Appendix \ref{Appendix_Explicit_Z} were derived starting from Eq. (\ref{Z_Result_Calculable}).

However, it is also expedient to present Eq. (\ref{Z_Result_General}) in such a way that the $\varepsilon$-poles and logarithms will be on the same footing. For this purpose, we note that in the integral over $t$ it is possible to make a small shift of the upper limit, $1\to 1-0$ by inserting the factor $\theta(1-0-t)$, where $\theta(x)=1$ for $x>0$ and $\theta(x)=0$ for $x<0$ is a step function. After that, in Eq. (\ref{T_Exponent}) for any $i$ we obtain $t_i<1$. From the other side, the argument of the exponential function containing $\ln\Lambda/\mu$ in Eq. (\ref{Z_Result_General}) may equivalently be rewritten as

\begin{equation}
\ln\frac{\Lambda}{\mu} \int\limits_0^{1+0} dt\, \delta(t-1)\Big( \frac{1}{t}\sum\limits_{i=1}^n \beta_i(t\alpha)\frac{\partial}{\partial\alpha_i} - \gamma(t\alpha)\Big).
\end{equation}

\noindent
Taking into account that (after the above mentioned shift of the upper limit) all $t_i$ inside the $T$-ordered exponential are less than 1, it is possible to combine both exponential functions in Eq. (\ref{Z_Result_General}) into a single ordered exponential,

\begin{eqnarray}\label{Z_TExponent}
&& Z\Big(\alpha,\varepsilon^{-1},\ln\frac{\Lambda}{\mu}\Big) = T \exp \bigg\{\int\limits_0^{1+0} dt \Big(\theta(1-0-t) \frac{1}{t\varepsilon} + \delta(t-1)\ln\frac{\Lambda}{\mu}\Big)\nonumber\\
&&\qquad\qquad\qquad\qquad\qquad\qquad\qquad\qquad\qquad\qquad \times\Big( \frac{1}{t}\sum\limits_{i=1}^n \beta_i(t\alpha)\frac{\partial}{\partial\alpha_i} - \gamma(t\alpha)\Big)\bigg\}\cdot 1.\qquad
\end{eqnarray}

\noindent
We see that $\varepsilon^{-1}$ and $\ln\Lambda/\mu$ enter this expression almost in the same way. This is important because, on the one hand, these two values are similar, but on the other hand, they enter the renormalization constants differently. In Eq. (\ref{Z_TExponent}) this difference appears because $\varepsilon^{-1}$ is ``smeared'' over the interval $[0,1)$, while $\ln\Lambda/\mu$ ``lives'' on its upper border.

Note that for theories with a single coupling in the MS-like schemes the expressions (\ref{Z_Result_Calculable}) and (\ref{Z_TExponent}) can be rewritten in a simpler form \cite{Meshcheriakov:2024qwj}

\begin{equation}\label{Z_Result_Single_Coupling}
Z\Big(\alpha,\varepsilon^{-1},\ln\frac{\Lambda}{\mu}\Big) = \exp\bigg\{\ln\frac{\Lambda}{\mu}\Big(\beta(\alpha)\frac{\partial}{\partial\alpha} - \gamma(\alpha)\Big)\bigg\}
\exp\bigg\{\int\limits_0^\alpha \frac{d\alpha\,\gamma(\alpha)}{\beta(\alpha)-\alpha\varepsilon}\bigg\}.
\end{equation}

\noindent
The derivation of this expression from Eq. (\ref{Z_Result_Calculable}) is presented in Appendix \ref{Appendix_Single_Coupling}.

It is also worth noting that the expression for $\ln Z$ is linear in the anomalous dimension and, therefore, simpler than $Z$. For the regularization described in Sect. \ref{Section_Regularization} we present it in Appendix \ref{Appendix_Ln_Z}. In the particular cases of the standard dimensional technique and of the cutoff-type regularization it is essentially simplified and can be written as

\begin{eqnarray}\label{Ln_Z_Result_DR}
&& \ln Z(\alpha,\varepsilon^{-1}) = - \Big(1-\int\limits^\wedge \frac{dt}{t^2\varepsilon}\sum\limits_{i=1}^n \beta_i(t\alpha)\frac{\partial}{\partial\alpha_i}\Big)^{-1} \int\limits^\wedge \frac{dt}{t\varepsilon}\,\gamma(t\alpha)\,\bigg|_{t=1};\qquad\\
\label{Ln_Z_Result_CutOff}
&& \ln Z(\alpha,\ln\Lambda/\mu) = - \int\limits_0^{\ln\Lambda/\mu} dt\,\exp\Big\{t \sum\limits_{i=1}^n \beta_i(\alpha)\frac{\partial}{\partial\alpha_i}\Big\}\, \gamma(\alpha),
\end{eqnarray}

\noindent
respectively. These equations are also very convenient for carrying out the explicit calculations in the lowest loops. However, the general expression (\ref{Ln_Z_Result}) is not so simple. Unlike Eq. (\ref{Z_TExponent}),  $\varepsilon$-poles and $\ln\Lambda/\mu$ enter it in a quite different way.

\section{Explicit expressions for renormalization constants in the three-loop approximation}
\hspace*{\parindent}\label{Section_Explicit_Result}

Let us present some explicit expressions in the lowest loops following from the general equations (\ref{Z_Result_Calculable}) and (\ref{Z_TExponent}). First, we consider a theory with two couplings, which will be denoted by $\alpha_1$ and $\alpha_2$. The perturbative expansions of RGFs in this case can be presented in the form

\begin{eqnarray}
&&\hspace*{-5mm} \gamma(\alpha_1,\alpha_2) = \sum\limits_{p,q=0;\ p+q\ne 0}^\infty \gamma_{pq}\alpha_1^p\alpha_2^q = \Big(\gamma_{10}\alpha_1 + \gamma_{01}\alpha_2\Big) + \Big(\gamma_{20}\alpha_1^2 + \gamma_{11}\alpha_1\alpha_2 + \gamma_{02}\alpha_2^2\Big)
+ \ldots;\nonumber\\
&&\hspace*{-5mm} \frac{\beta_1(\alpha_1,\alpha_2)}{\alpha_1} = \sum\limits_{p,q=0;\ p+q\ne 0}^\infty \beta_{1,pq} \alpha_1^p\alpha_2^q\nonumber\\
&&\hspace*{-5mm}\qquad\qquad\qquad\qquad\qquad = \Big(\beta_{1,10}\alpha_1 + \beta_{1,01}\alpha_2\Big) + \Big(\beta_{1,20}\alpha_1^2 + \beta_{1,11}\alpha_1\alpha_2 + \beta_{1,02}\alpha_2^2\Big) + \ldots;\qquad\nonumber\\
&&\hspace*{-5mm} \frac{\beta_2(\alpha_1,\alpha_2)}{\alpha_2} = \sum\limits_{p,q=0;\ p+q\ne 0}^\infty \beta_{2,pq} \alpha_1^p\alpha_2^q\nonumber\\
&&\hspace*{-5mm}\qquad\qquad\qquad\qquad\qquad = \Big(\beta_{2,10}\alpha_1 + \beta_{2,01}\alpha_2\Big) + \Big(\beta_{2,20}\alpha_1^2 + \beta_{2,11}\alpha_1\alpha_2 + \beta_{2,02}\alpha_2^2\Big) + \ldots\qquad
\end{eqnarray}

The general expression for the renormalization constant $Z$ in the three-loop approximation for dimensional regularization (or reduction) derived by expanding Eq. (\ref{Z_Result_Calculable}) in powers of $\alpha_i$ and setting $\Lambda=\mu$ is given by

\begin{eqnarray}\label{Z_Three-Loop_DR}
&&\hspace*{-5mm} Z = 1 - \frac{1}{\varepsilon}\Big[\alpha_1 \gamma_{10} +\alpha_2\gamma_{01}\Big]\nonumber\\
&&\hspace*{-5mm} + \frac{1}{2\varepsilon^2}\Big[\alpha_1^2\Big(\gamma_{10}^2 -\beta_{1,10} \gamma_{10}\Big)
+\alpha_1\alpha_2\Big(2\gamma_{10}\gamma_{01} - \beta_{1,01}\gamma_{10} -\beta_{2,10} \gamma_{01}\Big)
+ \alpha_2^2\Big(\gamma_{01}^2 -\beta_{2,01} \gamma_{01} \Big)\Big]\nonumber\\
&&\hspace*{-5mm} - \frac{1}{2\varepsilon}\Big[\alpha_1^2 \gamma_{20} +\alpha_1\alpha_2 \gamma_{11} +\alpha_2^2\gamma_{02}\Big]\nonumber\\
&&\hspace*{-5mm} - \frac{1}{6\varepsilon^3}\Big[\alpha_1^3 \Big(\gamma_{10}^3 -3\beta_{1,10}\gamma_{10}^2 + 2\beta_{1,10}^2\gamma_{10} \Big)
+\alpha_1^2\alpha_2 \Big(3\gamma_{10}^2 \gamma_{01} -3\beta_{1,01}\gamma_{10}^2 -3(\beta_{2,10} +\beta_{1,10})   \nonumber\\
&&\hspace*{-5mm}\qquad \times \gamma_{10}\gamma_{01} + (3\beta_{1,01}\beta_{1,10} +\beta_{1,01}\beta_{2,10})\gamma_{10} +(\beta_{2,10}^2 + \beta_{1,10}\beta_{2,10})\gamma_{01} \Big)
+\alpha_1\alpha_2^2\Big(3\gamma_{01}^2\gamma_{10} \nonumber\\
&&\hspace*{-5mm}\qquad -3\beta_{2,10}\gamma_{01}^2  -3(\beta_{1,01}+\beta_{2,01})\gamma_{01}\gamma_{10}
+ (\beta_{1,01}\beta_{2,10}+3\beta_{2,01}\beta_{2,10})\gamma_{01} +(\beta_{1,01}^2+\beta_{1,01}\vphantom{\Big(}\nonumber\\
&&\hspace*{-5mm}\qquad \times\beta_{2,01})\gamma_{10}\Big) +\alpha_2^3 \Big(\gamma_{01}^3 -3\beta_{2,01}\gamma_{01}^2 + 2\beta_{2,01}^2\gamma_{01}\Big)
\Big]\nonumber\\
&&\hspace*{-5mm} + \frac{1}{6\varepsilon^2}\Big[\alpha_1^3\Big(3\gamma_{10}\gamma_{20}-2\beta_{1,10}\gamma_{20}-2\beta_{1,20}\gamma_{10}\Big)
+\alpha_1^2\alpha_2\Big(3\gamma_{01}\gamma_{20} + 3\gamma_{10}\gamma_{11} -(\beta_{1,10}+\beta_{2,10})\nonumber\\
&&\hspace*{-5mm}\qquad \times\gamma_{11} -2\beta_{1,01}\gamma_{20}-2\beta_{1,11}\gamma_{10}-2\beta_{2,20}\gamma_{01}
\Big) + \alpha_1\alpha_2^2\Big(3\gamma_{02}\gamma_{10} +3\gamma_{01} \gamma_{11} -(\beta_{1,01}+\beta_{2,01})\nonumber\\
&&\hspace*{-5mm}\qquad\times\gamma_{11}-2\beta_{2,10}\gamma_{02}
-2\beta_{2,11}\gamma_{01}-2\beta_{1,02}\gamma_{10}\Big)
+ \alpha_2^3\Big(3\gamma_{01}\gamma_{02} -2\beta_{2,01}\gamma_{02} -2\beta_{2,02}\gamma_{01}\Big)
\Big]\nonumber\\
&&\hspace*{-5mm} - \frac{1}{3\varepsilon}\Big[\alpha_1^3 \gamma_{30} +\alpha_1^2\alpha_2 \gamma_{21} +\alpha_1\alpha_2^2 \gamma_{12} +\alpha_2^3\gamma_{03}\Big]
+ O(\alpha^4).
\end{eqnarray}

For the regularizations of the cutoff type the corresponding expression again follows from Eq. (\ref{Z_Result_Calculable}). However, in this case it is necessary to omit all $\varepsilon$-poles and keep powers of $\ln\Lambda/\mu$. The result in the three-loop approximation can be written in the form

\begin{eqnarray}\label{Z_Three-Loop_CutOff}
&&\hspace*{-5mm} Z = 1 - \ln\frac{\Lambda}{\mu} \Big[\alpha_1 \gamma_{10} +\alpha_2\gamma_{01}\Big]\nonumber\\
&&\hspace*{-5mm} + \frac{1}{2}\ln^2\frac{\Lambda}{\mu}\Big[\alpha_1^2\Big(\gamma_{10}^2 -\beta_{1,10} \gamma_{10}\Big)
+\alpha_1\alpha_2\Big(2\gamma_{10}\gamma_{01} - \beta_{1,01}\gamma_{10} -\beta_{2,10} \gamma_{01}\Big)
+ \alpha_2^2\Big(\gamma_{01}^2 -\beta_{2,01} \gamma_{01} \Big)\Big]\nonumber\\
&&\hspace*{-5mm} - \ln\frac{\Lambda}{\mu}\Big[\alpha_1^2 \gamma_{20} +\alpha_1\alpha_2 \gamma_{11} +\alpha_2^2\gamma_{02}\Big]\nonumber\\
&&\hspace*{-5mm} - \frac{1}{6}\ln^3\frac{\Lambda}{\mu}\Big[\alpha_1^3 \Big(\gamma_{10}^3 -3\beta_{1,10}\gamma_{10}^2 + 2\beta_{1,10}^2\gamma_{10}
\Big)
+\alpha_1^2\alpha_2 \Big(3\gamma_{10}^2 \gamma_{01} -3\beta_{1,01}\gamma_{10}^2 -3(\beta_{2,10} +\beta_{1,10})   \nonumber\\
&&\hspace*{-5mm}\qquad \times \gamma_{10}\gamma_{01}
+ (3\beta_{1,01}\beta_{1,10} +\beta_{1,01}\beta_{2,10})\gamma_{10} +(\beta_{2,10}^2 + \beta_{1,10}\beta_{2,10})\gamma_{01} \Big)
+\alpha_1\alpha_2^2\Big(3\gamma_{01}^2\gamma_{10} \nonumber\\
&&\hspace*{-5mm}\qquad -3\beta_{2,10}\gamma_{01}^2  -3(\beta_{1,01}+\beta_{2,01})\gamma_{01}\gamma_{10}
+ (\beta_{1,01}\beta_{2,10}+3\beta_{2,01}\beta_{2,10})\gamma_{01} +(\beta_{1,01}^2+\beta_{1,01}\vphantom{\Big(}\nonumber\\
&&\hspace*{-5mm}\qquad \times\beta_{2,01})\gamma_{10}\Big) +\alpha_2^3 \Big(\gamma_{01}^3 -3\beta_{2,01}\gamma_{01}^2 + 2\beta_{2,01}^2\gamma_{01}\Big)
\Big]\nonumber\\
&&\hspace*{-5mm} + \frac{1}{2}\ln^2\frac{\Lambda}{\mu}\Big[\alpha_1^3\Big(2\gamma_{10}\gamma_{20}-2\beta_{1,10}\gamma_{20}-\beta_{1,20}\gamma_{10}\Big)
+\alpha_1^2\alpha_2\Big(2\gamma_{01}\gamma_{20} + 2\gamma_{10}\gamma_{11} -(\beta_{1,10}+\beta_{2,10})\nonumber\\
&&\hspace*{-5mm}\qquad \times\gamma_{11} -2\beta_{1,01}\gamma_{20}-\beta_{1,11}\gamma_{10}-\beta_{2,20}\gamma_{01}\Big)
+ \alpha_1\alpha_2^2\Big(2\gamma_{02}\gamma_{10} +2\gamma_{01} \gamma_{11} -(\beta_{1,01}+\beta_{2,01})\nonumber\\
&&\hspace*{-5mm}\qquad\times\gamma_{11}-2\beta_{2,10}\gamma_{02}
-\beta_{2,11}\gamma_{01}-\beta_{1,02}\gamma_{10}\Big)
+ \alpha_2^3\Big(2\gamma_{01}\gamma_{02} -2\beta_{2,01}\gamma_{02} -\beta_{2,02}\gamma_{01}\Big)
\Big]\nonumber\\
&&\hspace*{-5mm} - \ln\frac{\Lambda}{\mu}\Big[\alpha_1^3 \gamma_{30} +\alpha_1^2\alpha_2 \gamma_{21} +\alpha_1\alpha_2^2 \gamma_{12} +\alpha_2^3\gamma_{03}\Big]
+ O(\alpha^4).
\end{eqnarray}

The (rather large) results containing both $\varepsilon$-poles, logarithms, and mixed terms in the three-loop approximation for MS-like renormalization prescriptions are presented in Appendix \ref{Appendix_Explicit_Z} for theories with two and three couplings. For a theory with two couplings we also present the expression for $\ln Z$ in the four-loop approximation. Certainly, the results for the regularizations of the dimensional and cutoff types can easily be obtained from them either by setting $\Lambda=\mu$, or by omitting $\varepsilon$-poles, respectively. However, it is always necessary to remember that changing the regularization we also change the coefficients of RGFs (starting from the three-loop approximation for the gauge $\beta$-functions or from the two-loop approximation for the anomalous dimensions).

\section{An example: two-loop renormalization of the matter superfields in ${\cal N}=1$ SQCD+SQED}
\hspace*{\parindent}\label{Section_Gamma}

As an example which illustrates the technique of calculation and allows to verify the correctness of general equations for the renormalization constants in the simplest nontrivial case, we consider ${\cal N}=1$ supersymmetric chromodynamics interacting with supersymmetric electrodynamics (SQCD+SQED) in the massless limit. It is convenient to write its action in terms of the superfields, because in this case supersymmetry becomes manifest both at the tree and quantum levels,

\begin{eqnarray}
&& S = \frac{1}{2g_0^2} \mbox{Re}\,\mbox{tr} \int d^4x\,d^2\theta\, W^a W_a + \frac{1}{2e_0^2} \mbox{Re}\,\int d^4x\,d^2\theta\, \bm{W}^a \bm{W}_a\nonumber\\
&&\qquad\qquad\qquad\qquad\qquad  + \sum\limits_{\mbox{\scriptsize a}=1}^{N_f}\frac{1}{4}\int d^4x\,d^4\theta\,\Big(\phi_{\mbox{\scriptsize a}}^+ e^{2V+2\bm{V}}\phi_{\mbox{\scriptsize a}} + \widetilde\phi_{\mbox{\scriptsize a}}^+ e^{-2V^T-2\bm{V}}\widetilde\phi_{\mbox{\scriptsize a}}\Big).\qquad
\end{eqnarray}

\noindent
This theory is invariant under the gauge transformations of the group $G\times U(1)$, where the group $G$ is simple, e.g., $G=SU(3)$ as for usual QCD. Therefore, the theory contains two (bare) gauge couplings $\alpha_{s0}\equiv g_{0}^2/4\pi$ and $\alpha_0=e_0^2/4\pi$. The former one corresponds to the (in general, non-Abelian) subgroup $G$, while the latter is the analog of the electromagnetic gauge coupling corresponding to the subgroup $U(1)$. The non-Abelian and Abelian gauge superfields are denoted by $V$ and $\bm{V}$ with the gauge field strengths $W_a=\bar{D}^2(e^{-2V}D_a e^{2V})/8$ and $\bm{W}_a =\bar{D}^2 D_a \bm{V}/4$, respectively. The theory contains $N_f$ flavors composed of the chiral matter superfields $(\phi_{\mbox{\scriptsize a}}, \widetilde\phi_{\mbox{\scriptsize a}})$ which lie in the mutually conjugate representations $R$ and $\xbar{R}$ of the group $G$ and have opposite $U(1)$ charges. For simplicity, we will assume that the representation $R$ is irreducible.

We will calculate quantum corrections to the two-point Green function of the matter superfields

\begin{equation}
\Gamma^{(2)}_\phi = \frac{1}{4} \int \frac{d^4p}{(2\pi)^4}\, d^4\theta\, \sum\limits_{\mbox{\scriptsize a}=1}^{N_f} \Big(\phi_{\mbox{\scriptsize a}}^+(-p,\theta) \phi_{\mbox{\scriptsize a}}(p,\theta) + \widetilde\phi_{\mbox{\scriptsize a}}^+(-p,\theta) \widetilde\phi_{\mbox{\scriptsize a}}(p,\theta)\Big)\, G(\alpha_0,\alpha_{s0},\varepsilon^{-1},\ln\Lambda/p)
\end{equation}

\noindent
using the version of the dimensional technique described in Sect.~\ref{Section_Regularization}. Nevertheless, it is necessary to remember that the dimensional regularization explicitly breaks supersymmetry \cite{Delbourgo:1974az}, so that we will in fact use dimensional reduction \cite{Siegel:1979wq}. This implies that the algebra of supersymmetric covariant derivatives is handled in four dimensions, while the integrals over loop momenta are calculated in $D=4-\varepsilon$ dimensions. We will use the dimensionless bare couplings $\alpha_0$ and $\alpha_{s0}$, so that the original (dimensionful in the regularized theory) ``electromagnetic'' and ``strong'' bare couplings should be presented as $\Lambda^\varepsilon\alpha_0$ and $\Lambda^\varepsilon\alpha_{s0}$, respectively.

\begin{figure}[h]
\includegraphics[scale=0.44]{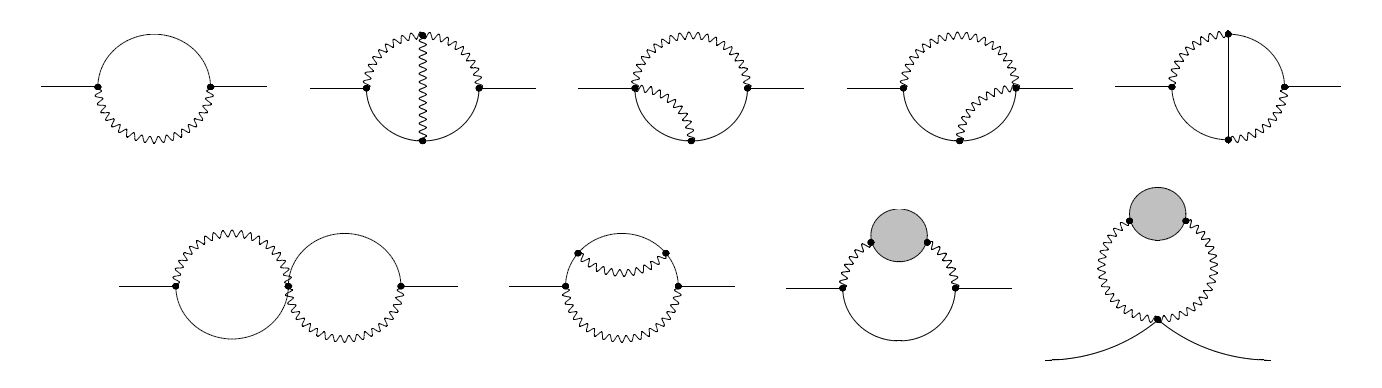}
\caption{The one- and two-loop supergraphs nontrivially contributing to the two-point Green function of the matter superfields in the Feynman gauge. The grey circles denote the insertions of the one-loop polarization operator.}\label{Figure_Supergraphs}
\end{figure}

It is convenient to make the calculations using the supersymmetric analog of the Feynman gauge, $\xi=1$. The one- and two-loop supergraphs contributing to the two-point Green function of the matter superfields and surviving in this gauge\footnote{The whole set of the one- and two-loop superdiagrams can be found in \cite{Kazantsev:2020kfl}.} are presented in Fig.~\ref{Figure_Supergraphs}.

The calculation made similarly to the case of using the higher covariant derivative regularization \cite{Kazantsev:2020kfl} gives the result for the function $G$ which can be written in the form

\begin{eqnarray}\label{LnG_Bare}
&&\hspace*{-7mm} \ln G = -8\pi\Big(\alpha_0 + \alpha_{s0} C(R)\Big)I_1 + 32\pi^2\Big(2\alpha_0^2 N_f\,\mbox{dim}\,R - 3\alpha_{s0}^2 C_2 C(R) + 2\alpha_{s0}^2 N_f T(R) C(R)\Big) I_2\nonumber\\
&&\hspace*{-7mm} + 32\pi^2 \Big(\alpha_0+\alpha_{s0} C(R)\Big)^2 \Big(2I_2 - I_1^2\Big) +\mbox{finite terms} + O(\alpha_0^3,\alpha_0^2\alpha_{s0},\alpha_0\alpha_{s0}^2,\alpha_{s0}^3).
\end{eqnarray}

\noindent
In our notation, $\mbox{dim}\,R = \delta_i^i$ is the dimension of the irreducible representation $R$, and the group Casimirs are defined by the equations

\begin{equation}
(T^A T^A)_i{}^j = C(R) \delta_i^j;\qquad \mbox{tr}(T^A T^B) = T(R)\delta^{AB};\qquad f^{ACD} f^{BCD} = C_2 \delta^{AB},
\end{equation}

\noindent
where $T^A$ are the generators of the group $G$ in the representation $R$, and $f^{ABC}$ are its structure constants, $[T^A,T^B] = if^{ABC} T^C$. The finite terms in Eq. (\ref{LnG_Bare}) are proportional to the external momentum $p$, and the Euclidean loop integrals $I_1$ and $I_2$ are defined by the equations

\begin{eqnarray}
&& I_1\equiv \Lambda^\varepsilon \int\frac{d^Dk}{(2\pi)^D}\,\frac{1}{k^2 (k+p)^2} = \frac{1}{(4\pi)^{2-\varepsilon/2}} \Big(\frac{\Lambda}{p}\Big)^\varepsilon \Gamma(\varepsilon/2) B(1-\varepsilon/2,1-\varepsilon/2);\\
&& I_2\equiv \Lambda^{2\varepsilon} \int\frac{d^Dk}{(2\pi)^D}\,\frac{d^Dl}{(2\pi)^D}\,\frac{1}{k^2 (k+p)^2 l^2 (k+l)^2}\nonumber\\
&&\qquad\quad\ \ = \frac{1}{(4\pi)^{4-\varepsilon}} \Big(\frac{\Lambda}{p}\Big)^{2\varepsilon} \frac{\Gamma(\varepsilon/2)\Gamma(\varepsilon)}{\Gamma(1+\varepsilon/2)} B(1-\varepsilon/2,1-\varepsilon/2) B(1-\varepsilon,1-\varepsilon/2).\qquad
\end{eqnarray}

For making the calculation, we will need the one-loop expressions for $\beta$-functions of the theory, which can be easily obtained by the standard methods,\footnote{For simplicity, we do not present here the corresponding (well-known) expressions for one-loop integrals defining the renormalization of the gauge couplings.}

\begin{eqnarray}\label{Beta_Expansions}
&& \frac{\beta(\alpha,\alpha_s)}{\alpha} = \frac{\alpha}{\pi} N_f\,\mbox{dim}\,R + O(\alpha^2,\alpha\alpha_s);\\
&& \frac{\beta_s(\alpha,\alpha_s)}{\alpha_s} = \frac{\alpha_s}{2\pi}\Big(2N_f T(R) - 3C_2\Big) + O(\alpha\alpha_s,\alpha_s^2).\qquad
\end{eqnarray}

\noindent
This in particular implies that the corresponding coefficients have the form

\begin{equation}\label{Beta_Coefficients}
\beta_{10} = \frac{1}{\pi} N_f\,\mbox{dim}\,R;\qquad \beta_{01} = 0;\qquad \beta_{s,10} = 0;\qquad \beta_{s,01} = \frac{1}{2\pi}\Big(2N_f T(R) - 3C_2\Big).
\end{equation}

\noindent
Taking these expressions into account, within the $D$-dimensional renormalization and using the $\overline{\mbox{DR}}$ scheme we present the bare couplings in the form

\begin{eqnarray}\label{Alpha_D_Renormalization}
&& \alpha_0 = \Big(\frac{\xbar{\mu}}{\Lambda}\Big)^{\varepsilon} \bm{\alpha} \Big(1+\frac{\bm{\alpha}}{\pi\varepsilon} N_f\,\mbox{dim}\,R + O(\bm{\alpha}^2,\bm{\alpha\alpha_s})\Big);\\
\label{AlphaS_D_Renormalization}
&& \alpha_{s0} = \Big(\frac{\xbar{\mu}}{\Lambda}\Big)^{\varepsilon} \bm{\alpha_s} \Big(1+\frac{\bm{\alpha_s}}{2\pi\varepsilon}\Big(2 N_f T(R) - 3C_2\Big) + O(\bm{\alpha\alpha_s},\bm{\alpha_s}^2)\Big),
\end{eqnarray}

\noindent
where $\bm{\alpha}$ and $\bm{\alpha_s}$ are the $D$-dimensional renormalized gauge couplings, and

\begin{equation}
\xbar\mu \equiv \frac{\mu \exp(\gamma/2)}{\sqrt{4\pi}}.
\end{equation}

\noindent
After substituting the expressions (\ref{Alpha_D_Renormalization}) and (\ref{AlphaS_D_Renormalization}) into Eq. (\ref{LnG_Bare}), we see that the dependence on the parameter $\Lambda$ disappears, while the remaining $\varepsilon$-poles can be removed by adding the expression

\begin{eqnarray}
&&\hspace*{-5mm} \ln\bm{Z} =  \frac{1}{\pi\varepsilon}\Big(\bm{\alpha} +\bm{\alpha_s} C(R)\Big) + \frac{1}{4\pi^2}\Big(2\bm{\alpha}^2 N_f\,\mbox{dim}\,R - 3\bm{\alpha_s}^2 C_2 C(R) + 2\bm{\alpha_s}^2 N_f T(R) C(R)\Big)
\nonumber\\
&&\hspace*{-5mm} \times\Big(\frac{1}{\varepsilon^2} - \frac{1}{2\varepsilon} \Big) - \frac{1}{4\pi^2\varepsilon}\Big(\bm{\alpha} +\bm{\alpha_s} C(R)\Big)^2 + O(\bm{\alpha}^3,\bm{\alpha}^2\bm{\alpha_s},\bm{\alpha}\bm{\alpha_s}^2,\bm{\alpha_s}^3).
\end{eqnarray}

\noindent
It is easy to verify that this expression exactly agrees with the (logarithm of the) expression (\ref{Z_Three-Loop_DR}) (or, equivalently, with Eq. (\ref{LnZ_Four-Loop_Two-Chages}) taken at $\Lambda=\mu$ for removing the logarithms) if the coefficients defining the anomalous dimension of the matter superfields are given by

\begin{eqnarray}\label{Gamma_Coefficients}
&& \gamma_{10} = -\frac{1}{\pi};\qquad\ \ \, \gamma_{01} = -\frac{1}{\pi} C(R);\qquad\ \ \, \gamma_{20} = \frac{1}{2\pi^2} \Big(1 + N_f\,\mbox{dim}\,R\Big);\qquad\nonumber\\
&& \gamma_{11} = \frac{1}{\pi^2} C(R);\qquad \gamma_{02} = \frac{1}{4\pi^2} \Big(2C(R)^2 - 3C_2 C(R) + 2 N_f T(R) C(R)\Big).
\end{eqnarray}

Next, we repeat the calculation using the four-dimensional renormalization technique in the $\overline{\mbox{DR}}$ scheme. In this case it is necessary to expand all quantum corrections in powers of $\varepsilon$ and include only $\varepsilon$-poles, powers of $\ln\Lambda/\,\xbar{\mu}$, and the mixed terms into the renormalization constants. In particular, in the one-loop approximation the renormalization of the gauge couplings is made according to the prescription

\begin{eqnarray}\label{Alpha_Renormalization}
&& \alpha_0 = \alpha + \frac{\alpha^2}{\pi} N_f\,\mbox{dim}\,R\, \Big(\frac{1}{\varepsilon} + \ln\frac{\Lambda}{\xbar{\mu}}\Big) + O(\alpha^3,\alpha^2\alpha_s);\\
\label{AlphaS_Renormalization}
&& \alpha_{s0} = \alpha_s + \frac{\alpha_s^2}{2\pi}\Big(2 N_f T(R) - 3C_2\Big)\Big(\frac{1}{\varepsilon} + \ln\frac{\Lambda}{\xbar{\mu}}\Big) + O(\alpha\alpha_s^2,\alpha_s^3)
\end{eqnarray}

\noindent
analogous to Eqs. (\ref{Alpha_D_Renormalization}) and (\ref{AlphaS_D_Renormalization}) within the $D$-dimensional technique.\footnote{The analogy becomes more clear if the renormalized couplings are expressed in terms of the bare ones.}

Rewriting the function $\ln G$ given by Eq. (\ref{LnG_Bare}) in terms of the four-dimensional renormalized gauge couplings $\alpha$ and $\alpha_s$ with the help of Eqs. (\ref{Alpha_Renormalization}) and (\ref{AlphaS_Renormalization}), we obtain

\begin{eqnarray}
&& \ln G = -8\pi\alpha\bigg[1+\frac{\alpha}{\pi}N_f\,\mbox{dim}\,R\Big(\frac{1}{\varepsilon}+\ln\frac{\Lambda}{\xbar{\mu}}\Big)\bigg] I_1
-8\pi\alpha_s C(R)\bigg[1+\frac{\alpha_s}{2\pi} \Big(2 N_f T(R) - 3C_2\Big)\qquad\nonumber\\
&&\qquad \times \Big(\frac{1}{\varepsilon}+\ln\frac{\Lambda}{\xbar{\mu}}\Big)\bigg] I_1
+ 32\pi^2\Big(2\alpha^2 N_f\,\mbox{dim}\,R - 3\alpha_s^2 C_2 C(R) + 2\alpha_s^2 N_f T(R) C(R)\Big) I_2\nonumber\\
&&\qquad + 32\pi^2 \Big(\alpha+\alpha_s C(R)\Big)^2 \Big(2I_2 - I_1^2\Big) +\mbox{finite terms} + O(\alpha^3,\alpha^2\alpha_s,\alpha\alpha_s^2,\alpha_s^3).
\end{eqnarray}

\noindent
Expanding this expression in a series in $\varepsilon$ and omitting all terms with its positive powers, we obtain that all divergences (including the ones containing powers of $\ln\Lambda$) can be eliminated by adding the expression

\begin{eqnarray}\label{LnZ_For_SQCD+SQED}
&&\hspace*{-5mm} \ln Z = \frac{1}{\pi}\Big(\alpha +\alpha_s C(R)\Big) \Big(\frac{1}{\varepsilon} + \ln\frac{\Lambda}{\xbar{\mu}}\Big) + \frac{1}{4\pi^2}\Big(2\alpha^2 N_f\,\mbox{dim}\,R - 3\alpha_s^2 C_2 C(R) + 2\alpha_s^2 N_f T(R) C(R)\Big)
\nonumber\\
&&\hspace*{-5mm}\times \bigg[\Big(\frac{1}{\varepsilon} + \ln\frac{\Lambda}{\xbar{\mu}}\Big)^2 - \Big(\frac{1}{2\varepsilon} + \ln\frac{\Lambda}{\xbar{\mu}}\Big) \bigg]
- \frac{1}{2\pi^2}\Big(\alpha +\alpha_s C(R)\Big)^2 \Big(\frac{1}{2\varepsilon} + \ln\frac{\Lambda}{\xbar{\mu}}\Big) + O(\alpha^3,\alpha^2\alpha_s,\alpha\alpha_s^2,\alpha_s^3),\nonumber\\
\end{eqnarray}

\noindent
which is certainly the logarithm of the renormalization constant for the chiral matter superfields. We see that this result has exactly the same structure as predicted by Eq. (\ref{LnZ_Four-Loop_Two-Chages}), which in the two-loop approximation for the theory under consideration (after the replacement $\mu\to\,\xbar{\mu}\,$) can be written in the form

\begin{eqnarray}\label{LnZ_General}
&&\hspace*{-7mm} \ln Z = - \Big(\frac{1}{\varepsilon}+\ln\frac{\Lambda}{\xbar{\mu}}\Big)\Big[\alpha\,\gamma_{10} +\alpha_s\,\gamma_{01}\Big] - \frac{1}{2}\Big(\frac{1}{\varepsilon}+\ln\frac{\Lambda}{\xbar{\mu}}\Big)^2\Big[
\alpha^2\,\beta_{10}\gamma_{10} +\alpha\alpha_s\Big(\beta_{01}\gamma_{10} + \beta_{s,10}\gamma_{01}\Big) \nonumber\\
&&\hspace*{-7mm} +\alpha_s^2\,\beta_{s,01}\gamma_{01}
\Big] - \Big(\frac{1}{2\varepsilon}+\ln\frac{\Lambda}{\xbar{\mu}}\Big)\Big[\alpha^2\,\gamma_{20} +\alpha\alpha_s\,\gamma_{11} +\alpha_s^2\,\gamma_{02}\Big]
+ O(\alpha^3,\alpha^2\alpha_s,\alpha\alpha_s^2,\alpha_s^3).
\end{eqnarray}

\noindent
In particular, we conclude that the coefficients defining the anomalous dimension are again given by the expressions (\ref{Gamma_Coefficients}), while the terms with higher poles and logarithms (i.e., proportional to $\varepsilon^{-2}$, $\varepsilon^{-1}\ln \Lambda/\,\xbar{\mu}$, and $\ln^2\Lambda/\,\xbar{\mu}$) entering Eq. (\ref{LnZ_For_SQCD+SQED}) really coincide with the prediction of Eq. (\ref{LnZ_General}) if we take Eq. (\ref{Beta_Coefficients}) into account. Certainly, this can be considered as a nontrivial test of the general expressions obtained in this paper and the calculation made for the theory under consideration.

Finally, we present the result for the anomalous dimension of the matter superfields (obtained for the $\overline{\mbox{DR}}$ renormalization prescription) following from Eq. (\ref{Gamma_Coefficients}),

\begin{eqnarray}\label{Gamma_Result}
&& \gamma(\alpha,\alpha_s) = -\frac{1}{\pi}\Big(\alpha+\alpha_s C(R)\Big) + \frac{1}{2\pi^2}\Big(\alpha+\alpha_s C(R)\Big)^2 + \frac{\alpha^2}{2\pi^2} N_f\,\mbox{dim}\,R \nonumber\\
&&\qquad\qquad\qquad\qquad\qquad - \frac{3\alpha_s^2}{4\pi^2} C_2 C(R) + \frac{\alpha_s^2}{2\pi^2} N_f T(R) C(R) + O(\alpha^3,\alpha^2\alpha_s,\alpha\alpha_s^2,\alpha_s^3).\qquad
\end{eqnarray}

\noindent
Comparing this expression with the $\overline{\mbox{DR}}$ results for the three-loop $\beta$-functions of the model calculated in \cite{Haneychuk:2025ehb}, we see that the NSVZ equations in this case \cite{Korneev:2021zdz,Kataev:2024amm} are not satisfied for the scheme dependent terms, while the scheme-independent consequences of the NSVZ equations (similar to those discussed in \cite{Kataev:2013csa}) are valid in exact agreement with the results of \cite{Jack:1996vg,Jack:1996cn,Jack:1998uj}.

\section{The three-loop renormalization constants for the $\varphi^4$-theory with two couplings}
\hspace*{\parindent}\label{Section_Kleinert}

For making another verification of general expressions derived in this paper, we consider the $\varphi^4$-theory described by the action

\begin{equation}\label{Varphi4_Two_Couplings_Lagrangian}
S = \int d^Dx\,\bigg\{ \frac{1}{2} \sum\limits_{a=1}^N \Big(\partial_\mu\varphi_a \partial^\mu\varphi_a - m_0^2 \varphi_a^2\Big) - \frac{16\pi^2 \Lambda^{\varepsilon}  g_{10}}{4!} \Big(\sum\limits_{a=1}^N \varphi_a^2 \Big)^2 - \frac{16\pi^2 \Lambda^{\varepsilon} g_{20}}{4!} \sum\limits_{a=1}^N \varphi_a^4\bigg\},
\end{equation}

\noindent
where the index $a$ range from 1 to $N$. Due to the factor $\Lambda^{\varepsilon}$, the bare couplings are dimensionless as in $D_0=4$. The theory contains two coupling constants $g_{10}$ and $g_{20}$.

The renormalization of $\varphi^4$-theory has been studied for a long time. For the particular case $g_{20}=0$ the four-loop RGFs were found in \cite{Kazakov:1979ik}. In the five-loop approximation the anomalous dimension of the scalar field has been calculated in \cite{Chetyrkin:1981jq,Chetyrkin:1981qh}. The $\beta$-function and the mass anomalous dimension in the five-loop approximation were found in \cite{Gorishnii:1983gp,Kazakov:1983dyk,Kazakov:1983ns}. The final five-loop results obtained after some corrections were given in \cite{Kleinert:1991rg}. The six-loop results were obtained in \cite{Kompaniets:2017yct}.

For the theory (\ref{Varphi4_Two_Couplings_Lagrangian}) the five-loop RGFs have been calculated in \cite{Kleinert:1994td}. In the six-loop approximation (for a more general theory) the RGFs were found in  \cite{Bednyakov:2021ojn}.

However, in this paper we are interested not in the RGFs, but in the renormalization constants. Various recursion relations between them in higher orders were checked in \cite{Kleinert:2001ax,Kastening:1996nj,Kastening:1997ah}. Here we will verify the general equations derived in this paper on the base of the explicit three-loop expressions for the renormalization constants presented in \cite{Kleinert:2001ax}.\footnote{Unfortunately, we did not manage to download the five-loop expressions for the renormalization constants from the link presented in \cite{Kleinert:2001ax} and are therefore unable to verify them.}

Taking into account that in our notations the bare couplings are dimensionless, we define the renormalization constants for the theory (\ref{Varphi4_Two_Couplings_Lagrangian}) by the equations

\begin{eqnarray}\label{Varphi4_Two_Couplings_Renormalization_Constants}
g_{10} = g_1 (Z_{g_1})^{-1};\qquad g_{20} = g_2 (Z_{g_2})^{-1};\qquad \varphi_a =\sqrt{Z_\varphi} \varphi_{a,R};\qquad  m_0 = \sqrt{Z_m} m.
\end{eqnarray}

\noindent
Here the bare coupling constants and the bare mass are marked by the subscript $0$, and $\varphi_R$ is the renormalized scalar field. The three-loop renormalization constants for the theory (\ref{Varphi4_Two_Couplings_Lagrangian}) can be found in \cite{Kleinert:2001ax}. They were calculated in the $\mbox{MS}$-like scheme which is obtained from the $\mbox{MS}$ scheme after the substitution

\begin{equation}
\mu \to \frac{\mu\,\exp(\gamma/2 +\varepsilon\zeta(2)/8)}{\sqrt{4\pi}},
\end{equation}

\noindent
where the Riemann $\zeta$-function is defined by the equation

\begin{equation}
\zeta(s) \equiv \sum\limits_{k=1}^\infty \frac{1}{k^s}.
\end{equation}

The three-loop expressions for the renormalization constants presented in \cite{Kleinert:2001ax} have the following form:

\begin{eqnarray}\label{Z_varphi}
&&\hspace*{-11mm} Z_\varphi = 1- \frac{1}{\varepsilon}\Big[\frac{g_1^2}{36}(2+N) + \frac{g_1 g_2}{6}  + \frac{g_2^2}{12}\Big]\nonumber\\
&&\hspace*{-7mm} - \frac{1}{\varepsilon^2}\Big[\frac{g_1^3}{162}(2+N)(8+N) + \frac{g_1^2 g_2}{18}(8+N) + \frac{g_1 g_2^2}{2} + \frac{g_2^3}{6}\Big]\nonumber\\
&&\hspace*{-7mm} + \frac{1}{\varepsilon}\Big[\frac{g_1^3}{648}(2+N)(8+N) + \frac{g_1^2 g_2}{72}(8+N) + \frac{g_1 g_2^2}{8} + \frac{g_2^3}{24}\Big] + O(g^4);\\
&& \vphantom{1}\nonumber\\
\label{Z_m}
&&\hspace*{-11mm} Z_{m} = 1 + \frac{1}{\varepsilon}\Big[ \frac{g_1}{3}(2+N) + g_2\Big]\nonumber\\
&&\hspace*{-7mm} + \frac{1}{\varepsilon^2}\Big[\frac{g_1^2}{9}(2+N)(5+N) + \frac{2g_1 g_2}{3}(5+N)  + 2 g_2^2 \Big]
- \frac{1}{\varepsilon}\Big[\frac{5g_1^2}{36}(2+N) + \frac{5 g_1 g_2}{6}  + \frac{5 g_2^2}{12}\Big]\nonumber\\
&&\hspace*{-7mm} + \frac{1}{\varepsilon^3}\Big[\frac{g_1^3}{27}(2+N)(5+N)(6+N) + \frac{g_1^2 g_2}{3}(5+N)(6+N) + \frac{14 g_1 g_2^2}{9}(8+N) + \frac{14g_2^3}{3}\Big]\nonumber\\
&&\hspace*{-7mm} - \frac{1}{\varepsilon^2}\Big[\frac{g_1^3}{324}(556+400N+61N^2) + \frac{g_1^2 g_2}{36}(278+61N) + \frac{g_1 g_2^2}{108}(982+35N) + \frac{113g_2^3}{36}\Big]\nonumber\\
&&\hspace*{-7mm} + \frac{1}{\varepsilon}\Big[\frac{g_1^3}{108}(74+47N+5N^2) + \frac{g_1^2 g_2}{12}(37+5N) + \frac{g_1 g_2^2}{72}(251+N) + \frac{7 g_2^3}{6}\Big] + O(g^4);\\
&&\vphantom{1}\nonumber\\
\label{Z_g1}
&&\hspace*{-11mm} (Z_{g_1})^{-1} = 1 + \frac{1}{\varepsilon}\Big[ \frac{g_1}{3}(8+N) + 2g_2\Big]\nonumber\\
&&\hspace*{-7mm} + \frac{1}{\varepsilon^2}\Big[\frac{g_1^2}{9}(8+N)^2 + g_1 g_2(12+N)  + 5 g_2^2 \Big]
- \frac{1}{\varepsilon}\Big[\frac{g_1^2}{6}(14+3N) + \frac{11 g_1 g_2}{3}  + \frac{5 g_2^2}{6}\Big]\nonumber\\
&&\hspace*{-7mm} + \frac{1}{\varepsilon^3}\Big[\frac{g_1^3}{27}(8+N)^3 + \frac{4g_1^2 g_2}{9}(116+21N+N^2) + \frac{2g_1 g_2^2}{9}(206+13N) + \frac{40g_2^3}{3}\Big]\nonumber\\
&&\hspace*{-7mm} - \frac{1}{\varepsilon^2}\Big[\frac{7g_1^3}{54}(8+N)(14+3N) + \frac{11g_1^2 g_2}{27}(94+11N) + \frac{g_1 g_2^2}{54}(1598+25N) + \frac{64g_2^3}{9}\Big]\nonumber\\
&&\hspace*{-7mm} + \frac{1}{\varepsilon}\Big[\frac{g_1^3}{324}(1480+461N+22N^2+1056\zeta(3)+240N\zeta(3))\nonumber\\
&&\hspace*{1mm} + \frac{g_1^2 g_2}{108}(1318+79N+768\zeta(3)) + \frac{g_1 g_2^2}{72}(642+N+192\zeta(3)) + \frac{7 g_2^3}{3}\Big]
+ O(g^4);\\
&&\vphantom{1}\nonumber\\
\label{Z_g2}
&&\hspace*{-11mm} (Z_{g_2})^{-1} = 1+ \frac{1}{\varepsilon}\Big[ 4 g_1 + 3 g_2\Big]\nonumber\\
&&\hspace*{-7mm} + \frac{1}{\varepsilon^2}\Big[\frac{2g_1^2}{3}(20+N) + 22 g_1 g_2 + 9 g_2^2 \Big]
- \frac{1}{\varepsilon}\Big[\frac{g_1^2}{18}(82+5N) + \frac{23 g_1 g_2}{3}  + \frac{17 g_2^2}{6}\Big]\nonumber\\
&&\hspace*{-7mm} + \frac{1}{\varepsilon^3}\Big[\frac{g_1^3}{27}(1120+136N+4N^2) + \frac{g_1^2 g_2}{3}(328+12N) + \frac{284g_1 g_2^2}{3} + 27 g_2^3\Big]\nonumber\\
&&\hspace*{-7mm} - \frac{1}{\varepsilon^2}\Big[\frac{g_1^3}{81}(2636+320N+5N^2) + \frac{g_1^2 g_2}{18}(1554+47N) + \frac{650g_1 g_2^2}{9} + \frac{119g_2^3}{6}\Big]\nonumber\\
&&\hspace*{-7mm} + \frac{1}{\varepsilon}\Big[\frac{g_1^3}{324}(6568+736N-13N^2+2688\zeta(3)+192N\zeta(3)) + \frac{g_1^2 g_2}{72}(1950+17N+1536\zeta(3))\nonumber\\
&&\hspace*{1mm} + \frac{g_1 g_2^2}{6}(131+96\zeta(3)) + \frac{g_2^3}{24}(145+96\zeta(3))\Big]
+ O(g^4).
\end{eqnarray}

To verify these expressions with the help of the general equations presented in this paper, first, we find the one- and two-loop coefficients of the anomalous dimensions and the $\beta$-functions. In our notations they are defined as

\begin{eqnarray}
&& \gamma_{\varphi}(g) \equiv \frac{d\ln Z_\varphi}{d\ln\mu}\bigg|_{g_0=\text{const}} =\sum\limits_{p,q=0;\ p+q\ne 0}^\infty \gamma_{\varphi,pq}\, g_1^p g_2^q; \qquad\nonumber\\
&& \gamma_{m}(g) \equiv \frac{d\ln Z_m}{d\ln\mu}\bigg|_{g_0=\text{const}} = \sum\limits_{p,q=0;\ p+q\ne 0}^\infty \gamma_{m,pq}\, g_1^p g_2^q;\nonumber\\
&& \beta_{g_i}(g) \equiv \frac{dg_i}{d\ln\mu}\bigg|_{g_0=\text{const}} = g_i \frac{d\ln Z_{g_i}}{d\ln\mu} = - g_i \frac{d\ln (Z_{g_i})^{-1}}{d\ln\mu} = g_i \sum\limits_{p,q=0;\ p+q\ne 0}^\infty \beta_{g_i,pq}\, g_1^p g_2^q.\qquad
\end{eqnarray}

\noindent
The coefficients of the RGFs perturbative expansions are related to the coefficients at simple $\varepsilon$-poles in the renormalization constants,

\begin{eqnarray}
&& Z = 1 - \sum\limits_{p,q=0;\ p+q\ne 0}^\infty\frac{g_1^p g_2^q}{(p+q)\varepsilon}\, \gamma_{pq}  + \mbox{higher poles};\qquad\nonumber\\
&& (Z_{g_i})^{-1} = 1 + \sum\limits_{p,q=0;\ p+q\ne 0}^\infty\frac{g_1^p g_2^q}{(p+q)\varepsilon}\, \beta_{g_i,pq}  + \mbox{higher poles},\qquad
\end{eqnarray}

\noindent
where $p+q=L$ is a number of loops. Then, the coefficients in the perturbative expansions of the RGFs under consideration following from Eqs. (\ref{Z_varphi}) --- (\ref{Z_g2}) are given by the expressions

\begin{eqnarray}
&&\hspace*{-7mm} \gamma_{\varphi,10} = 0;\qquad\qquad\ \ \gamma_{\varphi,01} = 0;\qquad \gamma_{\varphi,20} = \frac{2+N}{18};\qquad\qquad\ \gamma_{\varphi,11} = \frac{1}{3};\qquad\quad\ \
\gamma_{\varphi,02} = \frac{1}{6};\nonumber\\
&&\hspace*{-7mm} \gamma_{m,10} = -\frac{2+N}{3};\quad\  \gamma_{m,01} = -1;\quad\  \gamma_{m,20} = \frac{5(2+N)}{18};\qquad\ \ \gamma_{m,11} = \frac{5}{3};\qquad\quad\ \gamma_{m,02} = \frac{5}{6};\nonumber\\
&&\hspace*{-7mm} \beta_{g_1,10} = \frac{8+N}{3};\qquad \beta_{g_1,01} = 2;\qquad \beta_{g_1,20} = -\frac{14+3N}{3};\qquad \beta_{g_1,11} = -\frac{22}{3};\qquad \beta_{g_1,02} = -\frac{5}{3};\nonumber\\
&&\hspace*{-7mm} \beta_{g_2,10} = 4;\qquad\qquad\ \beta_{g_2,01} = 3;\qquad \beta_{g_2,20} = -\frac{82+5N}{9};\qquad \beta_{g_2,11} = -\frac{46}{3};\qquad \beta_{g_2,02} = -\frac{17}{3}.\nonumber\\
\end{eqnarray}

Substituting them into Eq. (\ref{Z_Three-Loop_DR}) we exactly reproduced all terms with higher poles in the renormalization constants (\ref{Z_varphi}) --- (\ref{Z_g2}). In particular, for constructing the expressions for $(Z_{g_i})^{-1}$, it is necessary to identify $\gamma_{pq}$ with $-\beta_{g_i,pq}$ because

\begin{equation}
\frac{d\ln (Z_{g_1})^{-1}}{d\ln\mu}\bigg|_{g_{0}=\text{const}} = - \frac{\beta_{g_1}}{g_1};\qquad \frac{d\ln (Z_{g_2})^{-1}}{d\ln\mu}\bigg|_{g_{0}=\text{const}} = - \frac{\beta_{g_2}}{g_2}.
\end{equation}

\noindent
Note that making the check we have detected a misprint in Eq. (15.30)\footnote{This equation corresponds to Eq. (\ref{Z_g1}) in this paper.} of Ref. \cite{Kleinert:2001ax}: the incorrect expression $4g_1^2 g_2 (116+16N+N^2)/9\varepsilon^3$ should be replaced with $4g_1^2 g_2(116+21N+N^2)/9\varepsilon^3$ as in Eq. (15.26) of Ref. \cite{Kleinert:2001ax}.

\section*{Conclusion}
\hspace*{\parindent}

In this paper we derive simple all-loop equations for the renormalization constants in theories with (an arbitrary number of) multiple couplings within the $\mbox{MS}$-like renormalization prescriptions. This can be done for both dimensional technique and regularizations of the cutoff type (like the higher covariant derivative regularization). However, it is most convenient to write down the result (given by Eqs. (\ref{Z_Result_Calculable}) and (\ref{Z_TExponent})) for the modification of the dimensional technique with $\Lambda\ne \mu$ because it allows reproducing the renormalization constants for both these types of the regularizations. Moreover, simultaneous presence of $\varepsilon$-poles and logarithms allows to trace the differences between coefficients at various powers of $\varepsilon^{-1}$ and $\ln\Lambda/\mu$. Note that in Eq. (\ref{Z_TExponent}) $\varepsilon^{-1}$ and $\ln\Lambda/\mu$ enter almost on the same footing. This is important because, from the one hand these two values are evidently analogous, but, from the other hand, enter quite differently into the renormalization constants.

Eqs. (\ref{Z_Result_Calculable}) and (\ref{Z_TExponent}) relate all coefficients at higher $\varepsilon$-poles, higher logarithms, and (if exist) mixed terms to the coefficients of RGFs, namely, of the anomalous dimension and the $\beta$-functions, in any order of the perturbation theory. These relations are very useful for verifying the correctness of various multiloop calculations. In particular, in this paper we write down some explicit expressions for the renormalization constants of the multicharge theories in the lowest orders of the perturbation theory and demonstrate how they can be used for checking the calculations. For this purpose, using the dimensional technique with two dimensional parameters $\Lambda$ and $\mu$ for ${\cal N}=1$ SQCD+SQED, we have calculated the two-loop renormalization constant of chiral matter superfields in the $\overline{\mbox{DR}}$ scheme and verified that the result agrees with general equations derived in this paper. Moreover, we consider the $\varphi^4$-theory with two couplings for that the explicit expressions for the three-loop renormalization constants were presented in \cite{Kleinert:2001ax} and demonstrated that these expressions (after correcting a minor misprint in \cite{Kleinert:2001ax}) really satisfy the general equations derived in this paper.

The technique developed in this paper can be applied for verifications of other future multiloop calculations, which are of great importance, for instance, in investigating the stability of the electroweak vacuum in the Standard Model and its various modifications, see, e.g. \cite{Bednyakov:2025uur} and references therein. Multiloop calculations made with regularizations of various types can also be used for investigating some interesting features of quantum corrections in certain supersymmetric theories, for example, the quantum properties of $P=1/3\, Q$ theories \cite{Jack:1996qq} or possibilities for the reduction of couplings, see \cite{Heinemeyer:2019vbc} and references therein.

\appendix

\section*{Appendix}

\section{Relations between integer-dimensional and $D$-dimensional RGFs for the MS-like renormalization prescriptions}
\hspace*{\parindent}\label{Appendix_RGFs_Relations}

Here we briefly recall the derivation of Eq. (\ref{RGFs_Relations}) for theories with multiple couplings. As a starting point, it is necessary to differentiate the equation

\begin{equation}
\frac{1}{\alpha_{i0}} \Big(\frac{\mu}{\Lambda}\Big)^\varepsilon = \frac{1}{\bm{\alpha}_i} \bm{Z}_{\alpha_i}
\end{equation}

\noindent
with respect to $\ln\mu$ at fixed values of $\alpha_{i0}$ and, after that, set $\Lambda=\mu$. This produces the equation

\begin{equation}\label{BetaD_Equation}
\varepsilon = \sum\limits_{j=1}^n\bm{\beta}_j(\alpha,\varepsilon) \frac{\partial}{\partial \alpha_j}\ln \Big(\frac{Z_{\alpha_i}(\alpha,\varepsilon^{-1})}{\alpha_i}\Big)
= -\frac{\bm{\beta}_i(\alpha,\varepsilon)}{\alpha_i} + \sum\limits_{j=1}^n\bm{\beta}_j(\alpha,\varepsilon) \frac{\partial}{\partial \alpha_j}\ln Z_{\alpha_i}(\alpha,\varepsilon^{-1}).
\end{equation}

\noindent
(Note that for $\Lambda=\mu$ the $D$-dimensional renormalized couplings $\bm{\alpha}_i$ and the integer-dimensional renormalized couplings $\alpha_i$ coincide, and so do the renormalization constants $\bm{Z}$ and $Z$. This implies that, after setting $\Lambda=\mu$, it is not necessary to write these values in bold.)

Similar equations for the four-/integer-dimensional $\beta$-functions are obtained with the help of the chain rule for the derivative with respect to $\ln\mu$ (also taken at fixed values of $\alpha_{i0}$),

\begin{equation}\label{Beta4_equation}
\beta_i(\alpha) = \alpha_i\frac{d\ln Z_{\alpha_i}}{d\ln\mu} = \alpha_i\sum\limits_{j=1}^n \frac{\partial\ln Z_{\alpha_i}}{\partial\alpha_j} \beta_j(\alpha) + \alpha_i\frac{\partial\ln Z_{\alpha_i}}{\partial\ln\mu}.
\end{equation}

\noindent
Certainly, this equation should be valid for any value of $\Lambda$ and, in particular, for $\Lambda=\mu$.

Next, comparing Eqs. (\ref{AlphaD_Renormalization}) and (\ref{Alpha4_Renormalization}) we notice that the
expressions $\bm{Z}(\alpha(\Lambda/\mu)^\varepsilon,\varepsilon^{-1})$ and $Z(\alpha,\ln\Lambda/\mu,\varepsilon^{-1})$ are similar. However, in fact, they are different, because the former one contains positive powers of $\varepsilon$, while in the latter one such terms are absent. Nevertheless, in the MS-like schemes (when all $Z(\alpha,\varepsilon^{-1})$ contain only $\varepsilon$-poles) the terms without $\ln\Lambda/\mu$ and the terms proportional to the first power of $\ln\Lambda/\mu$ coincide because in the former expression they cannot be multiplied to the positive powers of $\varepsilon$. This implies that the partial derivative in Eq. (\ref{Beta4_equation}) can be written in the form \cite{Narison:1980ti}

\begin{equation}\label{Z_MS_Auxiliary_Identity}
\frac{\partial\ln Z_{\alpha_i}}{\partial\ln\mu}\bigg|_{\Lambda=\mu} = -\varepsilon \sum\limits_{j=1}^n \alpha_j \frac{\partial\ln Z_{\alpha_i}}{\partial\alpha_j}\bigg|_{\Lambda=\mu}.
\end{equation}

\noindent
Substituting this expression into Eq. (\ref{Beta4_equation}) taken at $\Lambda=\mu$ yields the equation

\begin{equation}
\frac{\beta_i(\alpha)}{\alpha_i} = \sum\limits_{j=1}^n \Big(\beta_j(\alpha)-\varepsilon \alpha_j\Big)\frac{\partial\ln Z_{\alpha_i}(\alpha,\varepsilon^{-1})}{\partial\alpha_j}.
\end{equation}

\noindent
Comparing it with Eq. (\ref{BetaD_Equation}), we conclude that the $D$- and integer-dimensional $\beta$-functions in the MS-like renormalization schemes satisfy the equalities

\begin{equation}
\bm\beta_i(\alpha,\varepsilon) = - \varepsilon\alpha_i + \beta_i(\alpha).
\end{equation}

Similarly, the $D$-dimensional anomalous dimension defined by Eq. (\ref{RGFsD_Definition}) in the MS-like scheme is given by the expression

\begin{equation}\label{GammaD_Equation}
\bm{\gamma}(\alpha,\varepsilon) = \sum\limits_{i=1}^n \bm{\beta}_i(\alpha,\varepsilon) \frac{\partial \ln Z(\alpha,\varepsilon^{-1})}{\partial\alpha_i} = \sum\limits_{i=1}^n \Big(\beta_i(\alpha) - \varepsilon\alpha_i\Big)\frac{\partial \ln Z(\alpha,\varepsilon^{-1})}{\partial\alpha_i},
\end{equation}

\noindent
where we again assume that $\Lambda=\mu$. On the other side, applying the chain rule to the derivative with respect to $\ln\mu$ in the second equation of (\ref{RGFs4_Definition}), we present the integer-dimensional anomalous dimension in the form

\begin{equation}\label{Gamma4_Equation}
\gamma(\alpha) = \Big(\sum\limits_{i=1}^n \beta_i(\alpha) \frac{\partial\ln Z}{\partial\alpha_i} + \frac{\partial\ln Z}{\partial \ln\mu}\Big)\bigg|_{\Lambda=\mu} = \sum\limits_{i=1}^n \Big(\beta_i(\alpha) -\varepsilon\alpha_i\Big)\frac{\partial\ln Z(\alpha,\varepsilon^{-1})}{\partial\alpha_i},
\end{equation}

\noindent
where we also involve the equation analogous to Eq. (\ref{Z_MS_Auxiliary_Identity}) valid for the MS-like renormalization prescriptions. Comparing Eqs. (\ref{GammaD_Equation}) and (\ref{Gamma4_Equation}), we see that for the MS-like renormalization prescriptions both definitions of the anomalous dimension give the same result,

\begin{equation}
\bm{\gamma}(\alpha,\varepsilon) = \gamma(\alpha).
\end{equation}

\section{Explicit expressions for the renormalization constants in the lowest orders}
\hspace*{\parindent}\label{Appendix_Explicit_Z}

Although the general equations derived in this paper allow to construct easily the renormalization constants in any order of the perturbation theory, it is nevertheless reasonable to write down some explicit expressions in the lowest orders. For instance, these expressions can be used for checking the multiloop calculations. We will present them for the regularization described in Sect. \ref{Section_Regularization} because in this case both $\varepsilon$-poles and logarithms are present. This allows to trace the difference between the coefficients at higher $\varepsilon$-poles and higher logarithms. Moreover, the result for the standard dimensional regularization is obtained by setting $\Lambda=\mu$, while the result for the regularizations of the cutoff type is produced by removing all $\varepsilon$-poles (including the ones in the mixed terms).

First, we present the three-loop expression for a renormalization constant\footnote{Although the expressions in higher loops can easily be produced starting from Eq. (\ref{Z_Result_Calculable}), they are too large to be written here.} for a theory with two couplings,

\begin{eqnarray}\label{Z_Three-Loop_Two-Chages}
&&\hspace*{-5mm} Z = 1 - \Big(\frac{1}{\varepsilon}+\ln\frac{\Lambda}{\mu}\Big)\Big[\alpha_1 \gamma_{10} +\alpha_2\gamma_{01}\Big]\nonumber\\
&&\hspace*{-5mm} + \frac{1}{2}\Big(\frac{1}{\varepsilon}+\ln\frac{\Lambda}{\mu}\Big)^2\Big[\alpha_1^2\Big(\gamma_{10}^2 -\beta_{1,10} \gamma_{10}\Big)
+\alpha_1\alpha_2\Big(2\gamma_{10}\gamma_{01} - \beta_{1,01}\gamma_{10} -\beta_{2,10} \gamma_{01}\Big)
+ \alpha_2^2\Big(\gamma_{01}^2 \nonumber\\
&&\hspace*{-5mm}\qquad -\beta_{2,01} \gamma_{01} \Big)\Big]\nonumber\\
&&\hspace*{-5mm} - \Big(\frac{1}{2\varepsilon}+\ln\frac{\Lambda}{\mu}\Big)\Big[\alpha_1^2 \gamma_{20} +\alpha_1\alpha_2 \gamma_{11} +\alpha_2^2\gamma_{02}\Big]\nonumber\\
&&\hspace*{-5mm} - \frac{1}{6}\Big(\frac{1}{\varepsilon}+\ln\frac{\Lambda}{\mu}\Big)^3\Big[\alpha_1^3 \Big(\gamma_{10}^3 -3\beta_{1,10}\gamma_{10}^2 + 2\beta_{1,10}^2\gamma_{10} \Big)
+\alpha_1^2\alpha_2 \Big(3\gamma_{10}^2 \gamma_{01} -3\beta_{1,01}\gamma_{10}^2 -3(\beta_{2,10}    \nonumber\\
&&\hspace*{-5mm}\qquad +\beta_{1,10}) \gamma_{10}\gamma_{01} + (3\beta_{1,01}\beta_{1,10} +\beta_{1,01}\beta_{2,10})\gamma_{10} +(\beta_{1,10}\beta_{2,10} + \beta_{2,10}^2)\gamma_{01} \Big)
+\alpha_1\alpha_2^2\Big(3\gamma_{01}^2 \nonumber\\
&&\hspace*{-5mm}\qquad \times\gamma_{10} -3\beta_{2,10}\gamma_{01}^2  -3(\beta_{1,01}+\beta_{2,01})\gamma_{01}\gamma_{10}
+ (\beta_{1,01}\beta_{2,10}+3\beta_{2,01}\beta_{2,10})\gamma_{01} +(\beta_{1,01} \beta_{2,01}
\vphantom{\Big(}\nonumber\\
&&\hspace*{-5mm}\qquad + \beta_{1,01}^2)\gamma_{10}\Big) +\alpha_2^3 \Big(\gamma_{01}^3 -3\beta_{2,01}\gamma_{01}^2 + 2\beta_{2,01}^2\gamma_{01}\Big)
\Big]\nonumber\\
&&\hspace*{-5mm} + \Big(\frac{1}{2\varepsilon^2} + \frac{3}{2\varepsilon}\ln\frac{\Lambda}{\mu} + \ln^2\frac{\Lambda}{\mu}\Big)\Big[\alpha_1^3 \gamma_{10}\gamma_{20}
+\alpha_1^2\alpha_2\Big(\gamma_{01}\gamma_{20} + \gamma_{10}\gamma_{11}\Big) + \alpha_1\alpha_2^2\Big(\gamma_{02}\gamma_{10} +\gamma_{01} \gamma_{11} \Big)\nonumber\\
&&\hspace*{-5mm}\qquad + \alpha_2^3 \gamma_{01}\gamma_{02}
\Big]\nonumber\\
&&\hspace*{-5mm} - \Big(\frac{1}{6\varepsilon^2} + \frac{1}{2\varepsilon}\ln\frac{\Lambda}{\mu} +\frac{1}{2}\ln^2\frac{\Lambda}{\mu}\Big)\Big[2\alpha_1^3 \beta_{1,10}\gamma_{20}
+\alpha_1^2\alpha_2\Big( (\beta_{1,10}+\beta_{2,10}) \gamma_{11} +2\beta_{1,01}\gamma_{20}\Big) + \alpha_1\alpha_2^2\nonumber\\
&&\hspace*{-5mm}\qquad\times\Big( (\beta_{1,01}+\beta_{2,01}) \gamma_{11}+2\beta_{2,10}\gamma_{02}\Big)
+ 2\alpha_2^3 \beta_{2,01}\gamma_{02} \Big]\nonumber\\
&&\hspace*{-5mm} - \Big(\frac{1}{3\varepsilon^2} + \frac{1}{\varepsilon}\ln\frac{\Lambda}{\mu} + \frac{1}{2} \ln^2\frac{\Lambda}{\mu}\Big)\Big[\alpha_1^3 \beta_{1,20}\gamma_{10}
+\alpha_1^2\alpha_2\Big(  \beta_{1,11}\gamma_{10} + \beta_{2,20}\gamma_{01} \Big) + \alpha_1\alpha_2^2\Big(\beta_{2,11}\gamma_{01} \nonumber\\
&&\hspace*{-5mm}\qquad + \beta_{1,02}\gamma_{10}\Big)+ \alpha_2^3 \beta_{2,02}\gamma_{01} \Big]\nonumber\\
&&\hspace*{-5mm} - \Big(\frac{1}{3\varepsilon}+\ln\frac{\Lambda}{\mu}\Big)\Big[\alpha_1^3 \gamma_{30} +\alpha_1^2\alpha_2 \gamma_{21} +\alpha_1\alpha_2^2 \gamma_{12} +\alpha_2^3\gamma_{03}\Big]
+ O(\alpha^4).
\end{eqnarray}

\noindent
Then the standard MS scheme is produced for $\Lambda=\mu$, and the minimal subtractions of logarithms are obtained after removing all $\varepsilon$-poles. The results for these two particular cases are given by Eqs. (\ref{Z_Three-Loop_DR}) and (\ref{Z_Three-Loop_CutOff}), respectively.

In the four-loop approximation, we present the result for $\ln Z$, because it is linear in the anomalous dimension and, therefore, smaller than the expression for $Z$,\footnote{This expression has been derived from both Eq. (\ref{Z_Result_Calculable}) and Eq. (\ref{Ln_Z_Result}). The coincidence of the results has been checked.}

\begin{eqnarray}\label{LnZ_Four-Loop_Two-Chages}
&&\hspace*{-5mm} \ln Z = - \Big(\frac{1}{\varepsilon}+\ln\frac{\Lambda}{\mu}\Big)\Big[\alpha_1\,\gamma_{10} +\alpha_2\,\gamma_{01}\Big]\nonumber\\
&&\hspace*{-5mm} - \frac{1}{2}\Big(\frac{1}{\varepsilon}+\ln\frac{\Lambda}{\mu}\Big)^2\Big[
\alpha_1^2\,\beta_{1,10}\gamma_{10} +\alpha_1\alpha_2\Big(\beta_{1,01}\gamma_{10} + \beta_{2,10}\gamma_{01}\Big) +\alpha_2^2\,\beta_{2,01}\gamma_{01}
\Big]\nonumber\\
&&\hspace*{-5mm} - \Big(\frac{1}{2\varepsilon}+\ln\frac{\Lambda}{\mu}\Big)\Big[\alpha_1^2\,\gamma_{20} +\alpha_1\alpha_2\,\gamma_{11} +\alpha_2^2\,\gamma_{02}\Big]\nonumber\\
&&\hspace*{-5mm} - \frac{1}{6}\Big(\frac{1}{\varepsilon}+\ln\frac{\Lambda}{\mu}\Big)^3\Big[ 2\alpha_1^3\,\beta_{1,10}^2\gamma_{10}
+\alpha_1^2\alpha_2\Big((3\beta_{1,10}+\beta_{2,10})\beta_{1,01}\gamma_{10} +(\beta_{1,10}+\beta_{2,10}) \beta_{2,10}\gamma_{01} \Big)
\nonumber\\
&&\hspace*{-5mm}\qquad
+\alpha_1\alpha_2^2\Big((\beta_{1,01}+\beta_{2,01})\beta_{1,01}\gamma_{10}+(\beta_{1,01}+3\beta_{2,01})\beta_{2,10}\gamma_{01}\Big)
+2\alpha_2^3\,\beta_{2,01}^2\gamma_{01}\Big]\nonumber\\
&&\hspace*{-5mm} - \Big(\frac{1}{6\varepsilon^2} + \frac{1}{2\varepsilon}\ln\frac{\Lambda}{\mu} + \frac{1}{2}\ln^2\frac{\Lambda}{\mu}\Big)\Big[
2\alpha_1^3\,\beta_{1,10}\gamma_{20}
+\alpha_1^2\alpha_2\Big((\beta_{1,10} + \beta_{2,10})\gamma_{11} +2\beta_{1,01}\gamma_{20}\Big)
\nonumber\\
&&\hspace*{-5mm}\qquad
+\alpha_1 \alpha_2^2\Big((\beta_{1,01} + \beta_{2,01})\gamma_{11} +2\beta_{2,10}\gamma_{02}\Big)
+2\alpha_2^3\,\beta_{2,01}\gamma_{02}
\Big]\nonumber\\
&&\hspace*{-5mm} - \Big(\frac{1}{3\varepsilon^2} + \frac{1}{\varepsilon}\ln\frac{\Lambda}{\mu} + \frac{1}{2} \ln^2\frac{\Lambda}{\mu}\Big)\Big[
\alpha_1^3\,\beta_{1,20}\gamma_{10}
+\alpha_1^2\alpha_2\Big(\beta_{1,11}\gamma_{10} + \beta_{2,20}\gamma_{01}\Big)
+\alpha_1\alpha_2^2\Big(\beta_{1,02}\gamma_{10}
\nonumber\\
&&\hspace*{-5mm}\qquad
+\beta_{2,11}\gamma_{01}\Big)+\alpha_2^3\beta_{2,02}\gamma_{01} \Big]\nonumber\\
&&\hspace*{-5mm} - \Big(\frac{1}{3\varepsilon}+\ln\frac{\Lambda}{\mu}\Big)\Big[\alpha_1^3\,\gamma_{30} +\alpha_1^2\alpha_2\,\gamma_{21} +\alpha_1\alpha_2^2\,\gamma_{12} +\alpha_2^3\,\gamma_{03}\Big]
\nonumber\\
&&\hspace*{-5mm} - \frac{1}{24}\Big(\frac{1}{\varepsilon}+\ln\frac{\Lambda}{\mu}\Big)^4\Big[
6\alpha_1^4\,\beta_{1,10}^3\gamma_{10}
+\alpha_1^3\alpha_2\Big((\beta_{2,10}^2 +3\beta_{1,10}\beta_{2,10} +2\beta_{1,10}^2)\beta_{2,10}\gamma_{01}
+(\beta_{2,10}^2\nonumber\\
&&\hspace*{-5mm}\qquad +5\beta_{1,10}\beta_{2,10} +12 \beta_{1,10}^2)\beta_{1,01}\gamma_{10}\Big)
+\alpha_1^2\alpha_2^2\Big((3\beta_{1,01}\beta_{1,10}+4\beta_{1,10}\beta_{2,01}+4\beta_{1,01}\beta_{2,10}
\nonumber\\
&&\hspace*{-5mm}\qquad
+7\beta_{2,01}\beta_{2,10})\beta_{2,10} \gamma_{01}
+(3\beta_{2,01}\beta_{2,10}+4\beta_{1,10}\beta_{2,01}+4\beta_{1,01}\beta_{2,10}+7\beta_{1,01}\beta_{1,10})\beta_{1,01}
\vphantom{\Big(}\nonumber\\
&&\hspace*{-5mm}\qquad
\times \gamma_{10}\Big)
+\alpha_1\alpha_2^3\Big((\beta_{1,01}^2+3\beta_{1,01}\beta_{2,01}+2\beta_{2,01}^2)\beta_{1,01}\gamma_{10}
+(\beta_{1,01}^2+5\beta_{1,01}\beta_{2,01}+12\nonumber\\
&&\hspace*{-5mm}\qquad
\times\beta_{2,01}^2)\beta_{2,10}\gamma_{01}\Big)
+6\alpha_2^4\,\beta_{2,01}^3\gamma_{01}\Big]\nonumber\\
&&\hspace*{-5mm} - \frac{1}{24}\Big(\frac{1}{\varepsilon^3} + \frac{4}{\varepsilon^2}\ln\frac{\Lambda}{\mu} + \frac{6}{\varepsilon} \ln^2\frac{\Lambda}{\mu} + 4\ln^3\frac{\Lambda}{\mu}\Big)\Big[
6\alpha_1^4\,\beta_{1,10}^2\gamma_{20}
+\alpha_1^3\alpha_2\Big((2\beta_{2,10}+10\beta_{1,10})\beta_{1,01}
\vphantom{\Big(}\nonumber\\
&&\hspace*{-5mm}\qquad
\times\gamma_{20}+(\beta_{2,10}^2+3\beta_{1,10}\beta_{2,10}+2\beta_{1,10}^2)\gamma_{11}\Big)
+\alpha_1^2\alpha_2^2\Big(
(4\beta_{1,01}^2+2\beta_{1,01}\beta_{2,01})\gamma_{20}
+(4\beta_{2,10}^2
\vphantom{\Big(}\nonumber\\
&&\hspace*{-5mm}\qquad
+2\beta_{1,10}\beta_{2,10})\gamma_{02}
+(3\beta_{2,01}\beta_{2,10}
+3\beta_{1,01}\beta_{1,10}
+4\beta_{1,01}\beta_{2,10}
+2\beta_{1,10}\beta_{2,01})\gamma_{11}
\Big)+\alpha_1
\nonumber\\
&&\hspace*{-5mm}\qquad
\times\alpha_2^3 \Big(
(\beta_{1,01}^2 +3\beta_{1,01}\beta_{2,01} +2\beta_{2,01}^2)\gamma_{11}
+(2\beta_{1,01}+10\beta_{2,01})\beta_{2,10}\gamma_{02} \Big)
+6\alpha_2^4\,\beta_{2,01}^2\gamma_{02}
\Big]\nonumber\\
&&\hspace*{-5mm}
- \frac{1}{24\varepsilon^2}\Big(\frac{1}{\varepsilon} + 4\ln\frac{\Lambda}{\mu} \Big)\Big[
12\alpha_1^4\,\beta_{1,10}\beta_{1,20}\gamma_{10}
+\alpha_1^3\alpha_2\Big((2\beta_{1,11}\beta_{2,10}+3\beta_{1,01}\beta_{2,20}+9\beta_{1,01}\beta_{1,20}
\nonumber\\
&&\hspace*{-5mm}\qquad
+10\beta_{1,10}\beta_{1,11})\gamma_{10}
+(3\beta_{1,20}\beta_{2,10}+4\beta_{1,10}\beta_{2,20}+5\beta_{2,10}\beta_{2,20})\gamma_{01}
\Big)
+\alpha_1^2\alpha_2^2\Big(
(2\beta_{1,11}\beta_{2,01}
\nonumber\\
&&\hspace*{-5mm}\qquad
+3\beta_{1,01}\beta_{2,11}+4\beta_{1,02}\beta_{2,10}+7\beta_{1,01}\beta_{1,11}+8\beta_{1,02}\beta_{1,10})\gamma_{10}
+(2\beta_{1,10}\beta_{2,11}+3\beta_{1,11}\beta_{2,10}
\vphantom{\Big(}\nonumber\\
&&\hspace*{-5mm}\qquad
+4\beta_{1,01}\beta_{2,20}+7\beta_{2,10}\beta_{2,11}+8\beta_{2,01}\beta_{2,20})\gamma_{01}
\Big)
+\alpha_1\alpha_2^3\Big(
(2\beta_{1,01}\beta_{2,11}+3\beta_{1,02}\beta_{2,10}
\nonumber\\
&&\hspace*{-5mm}\qquad
+9\beta_{2,02}\beta_{2,10}+10\beta_{2,01}\beta_{2,11})\gamma_{01}
+(3\beta_{1,01}\beta_{2,02}+4\beta_{1,02}\beta_{2,01}+5\beta_{1,01}\beta_{1,02})\gamma_{10}
\Big)
+12\alpha_2^4\,\nonumber\\
&&\hspace*{-5mm}\qquad
\times\beta_{2,01}\beta_{2,02}\gamma_{01}
\Big]
\nonumber\\
&&\hspace*{-5mm}
- \frac{1}{2}\ln^2\frac{\Lambda}{\mu}\Big(\frac{1}{\varepsilon} +\frac{1}{3}\ln\frac{\Lambda}{\mu}\Big)\Big[
5\alpha_1^4\,\beta_{1,20}\beta_{1,10}\gamma_{10}
+\alpha_1^3\alpha_2\Big(
(\beta_{1,11}\beta_{2,10} +\beta_{1,01}\beta_{2,20} +4\beta_{1,11}\beta_{1,10}
\nonumber\\
&&\hspace*{-5mm}\qquad
+4\beta_{1,20}\beta_{1,01})\gamma_{10}
+(\beta_{2,10}\beta_{1,20}+2\beta_{2,20}\beta_{1,10}+2\beta_{2,20}\beta_{2,10})\gamma_{01}
\Big)
+\alpha_1^2\alpha_2^2\Big(
(\beta_{1,01}\beta_{2,11}
\nonumber\\
&&\hspace*{-5mm}\qquad
+\beta_{1,11}\beta_{2,01}+2\beta_{1,02}\beta_{2,10}+3\beta_{1,02}\beta_{1,10}+3\beta_{1,11}\beta_{1,01})\gamma_{10}
+(\beta_{2,11}\beta_{1,10}+\beta_{2,10}\beta_{1,11}
\vphantom{\Big(}\nonumber\\
&&\hspace*{-5mm}\qquad
+2\beta_{2,20}\beta_{1,01}+3\beta_{2,20}\beta_{2,01}+3\beta_{2,11}\beta_{2,10})\gamma_{01}
\Big)
+\alpha_1\alpha_2^3\Big(
(\beta_{2,10}\beta_{1,02}+\beta_{2,11}\beta_{1,01}+4
\vphantom{\Big(}\nonumber\\
&&\hspace*{-5mm}\qquad
\times \beta_{2,02}\beta_{2,10}+4\beta_{2,11}\beta_{2,01})\gamma_{01}
+(\beta_{1,01}\beta_{2,02}
+2\beta_{1,02}\beta_{1,01}
+2\beta_{1,02}\beta_{2,01})\gamma_{10}
\Big)
+5\alpha_2^4\,\nonumber\\
&&\hspace*{-5mm}\qquad
\times
\beta_{2,02}\beta_{2,01}\gamma_{01}
\Big]
\nonumber\\
&&\hspace*{-5mm} - \Big(\frac{1}{12\varepsilon^2} + \frac{1}{3\varepsilon}\ln\frac{\Lambda}{\mu} + \frac{1}{2}\ln^2\frac{\Lambda}{\mu} \Big)\Big[
3\alpha_1^4\,\beta_{1,10}\gamma_{30}
+ \alpha_1^3\alpha_2\Big((\beta_{2,10}+2\beta_{1,10})\gamma_{21}+3\beta_{1,01}\gamma_{30}\Big)
\nonumber\\
&&\hspace*{-5mm} \qquad
+\alpha_1^2\alpha_2^2\Big((\beta_{2,01}+2\beta_{1,01})\gamma_{21}+(\beta_{1,10}+2\beta_{2,10})\gamma_{12}\Big)
+\alpha_1\alpha_2^3\Big((\beta_{1,01}+2\beta_{2,01})\gamma_{12}
\nonumber\\
&&\hspace*{-5mm} \qquad
+3\beta_{2,10}\gamma_{03}\Big)
+3\alpha_2^4\,\beta_{2,01}\gamma_{03}
\Big]\nonumber\\
&&\hspace*{-5mm} - \frac{1}{2}\Big(\frac{1}{4\varepsilon^2} + \frac{1}{\varepsilon}\ln\frac{\Lambda}{\mu} + \ln^2\frac{\Lambda}{\mu} \Big)\Big[
2\alpha_1^4\,\beta_{1,20}\gamma_{20}
+\alpha_1^3\alpha_2\Big((\beta_{1,20}+\beta_{2,20})\gamma_{11}+2\beta_{1,11}\gamma_{20}\Big)
\nonumber\\
&&\hspace*{-5mm}\qquad
+\alpha_1^2\alpha_2^2\Big( 2\beta_{1,02}\gamma_{20} +2\beta_{2,20}\gamma_{02} +(\beta_{1,11}+\beta_{2,11})\gamma_{11}\Big)
+\alpha_1\alpha_2^3\Big((\beta_{1,02} + \beta_{2,02})\gamma_{11}
\nonumber\\
&&\hspace*{-5mm}\qquad
+2\beta_{2,11}\gamma_{02}\Big)
+2\alpha_2^4\,\beta_{2,02}\gamma_{02}
\Big]\nonumber\\
&&\hspace*{-5mm} - \Big(\frac{1}{4\varepsilon^2} + \frac{1}{\varepsilon}\ln\frac{\Lambda}{\mu} + \frac{1}{2}\ln^2\frac{\Lambda}{\mu} \Big)\Big[
\alpha_1^4\,\beta_{1,30}\gamma_{10}
+\alpha_1^3\alpha_2\Big(\beta_{2,30}\gamma_{01} +\beta_{1,21}\gamma_{10}\Big)
+\alpha_1^2\alpha_2^2\Big(\beta_{1,12}
\nonumber\\
&&\hspace*{-5mm}\qquad
\times\gamma_{10}+\beta_{2,21}\gamma_{01}\Big)
+\alpha_1\alpha_2^3\Big(\beta_{1,03}\gamma_{10}+\beta_{2,12}\gamma_{01}\Big)
+\alpha_2^4\,\beta_{2,03}\gamma_{01}
\Big]\nonumber\\
&&\hspace*{-5mm} - \Big(\frac{1}{4\varepsilon} + \ln\frac{\Lambda}{\mu} \Big)\Big[
\alpha_1^4\,\gamma_{40} + \alpha_1^3\alpha_2\,\gamma_{31} +\alpha_1^2\alpha_2^2\,\gamma_{22}
+\alpha_1\alpha_2^3\,\gamma_{13} +\alpha_2^4\,\gamma_{04}\Big] + O(\alpha^5).
\end{eqnarray}

For a theory with three couplings we use the notations

\begin{eqnarray}
&& \gamma(\alpha_1,\alpha_2,\alpha_3) = \sum\limits_{p,q,r=0;\,p+q+r\ne 0}^\infty \alpha_1^p \alpha_2^q \alpha_3^r\, \gamma_{pqr} = \alpha_1 \gamma_{100} + \alpha_2\gamma_{010} + \alpha_3\gamma_{001} + \ldots;\nonumber\\
&& \frac{\beta_k(\alpha_1,\alpha_2,\alpha_3)}{\alpha_k} = \sum\limits_{p,q,r=0;\,p+q+r\ne 0}^\infty \alpha_1^p \alpha_2^q \alpha_3^r\, \beta_{k,pqr} = \alpha_1 \beta_{k,100} + \alpha_2\beta_{k,010} + \alpha_3\beta_{k,001} + \ldots,\qquad
\end{eqnarray}

\noindent
where $k=1,2,3$. In this case the expression for a renormalization constant analogous to Eq. (\ref{Z_Three-Loop_Two-Chages}) has the form

\begin{eqnarray}\label{Z_Three-Loop_Three-Chages}
&&\hspace*{-5mm} Z = 1 - \Big(\frac{1}{\varepsilon}+\ln\frac{\Lambda}{\mu}\Big)\Big[\alpha_1 \gamma_{100} +\alpha_2\gamma_{010} +\alpha_3\gamma_{001}\Big]\nonumber\\
&&\hspace*{-5mm} + \frac{1}{2}\Big(\frac{1}{\varepsilon}+\ln\frac{\Lambda}{\mu}\Big)^2\Big[\alpha_1^2\Big(\gamma_{100}^2 -\beta_{1,100} \gamma_{100}\Big) + \alpha_2^2\Big(\gamma_{010}^2 -\beta_{2,010} \gamma_{010} \Big)
+ \alpha_3^2\Big(\gamma_{001}^2 -\beta_{3,001} \gamma_{001} \Big)\nonumber\\
&&\hspace{-5mm}\qquad
+\alpha_1\alpha_2\Big(2\gamma_{100}\gamma_{010} - \beta_{1,010}\gamma_{100} -\beta_{2,100} \gamma_{010}\Big)
+\alpha_1\alpha_3\Big(2\gamma_{100}\gamma_{001} - \beta_{1,001}\gamma_{100} -\beta_{3,100} \nonumber\\
&&\hspace{-5mm}\qquad
\times \gamma_{001}\Big) +\alpha_2\alpha_3\Big(2\gamma_{010}\gamma_{001} - \beta_{2,001}\gamma_{010} -\beta_{3,010} \gamma_{001}\Big)
\Big]\nonumber\\
&&\hspace*{-5mm} - \Big(\frac{1}{2\varepsilon}+\ln\frac{\Lambda}{\mu}\Big)\Big[\alpha_1^2 \gamma_{200} +\alpha_2^2\gamma_{020} +\alpha_3^2\gamma_{002} +\alpha_1\alpha_2 \gamma_{110} +\alpha_1\alpha_3 \gamma_{101} +\alpha_2\alpha_3 \gamma_{011}\Big]\nonumber\\
&&\hspace*{-5mm} - \frac{1}{6}\Big(\frac{1}{\varepsilon}+\ln\frac{\Lambda}{\mu}\Big)^3\Big[\alpha_1^3 \Big(\gamma_{100}^3 -3\beta_{1,100}\gamma_{100}^2 + 2\beta_{1,100}^2\gamma_{100} \Big)
+\alpha_2^3 \Big(\gamma_{010}^3 -3\beta_{2,010}\gamma_{010}^2 + 2\beta_{2,010}^2\nonumber\\
&&\hspace*{-5mm}\qquad \times \gamma_{010}\Big) +\alpha_3^3 \Big(\gamma_{001}^3 -3\beta_{3,001}\gamma_{001}^2 + 2\beta_{3,001}^2\gamma_{001}\Big)
\nonumber\\
&&\hspace*{-5mm}\qquad
+\alpha_1^2\alpha_2 \Big(3\gamma_{100}^2 \gamma_{010} -3\beta_{1,010}\gamma_{100}^2 -3 (\beta_{2,100} +\beta_{1,100}) \gamma_{100}\gamma_{010} + (3\beta_{1,010}\beta_{1,100} +\beta_{1,010}
\nonumber\\
&&\hspace*{-5mm}\qquad
\times \beta_{2,100})\gamma_{100} +(\beta_{1,100}\beta_{2,100} + \beta_{2,100}^2)\gamma_{010} \Big)
+\alpha_1\alpha_2^2\Big(3\gamma_{010}^2 \gamma_{100} -3\beta_{2,100}\gamma_{010}^2  -3(\beta_{1,010}
\nonumber\\
&&\hspace*{-5mm}\qquad
+\beta_{2,010})\gamma_{010}\gamma_{100} + (\beta_{1,010}\beta_{2,100} +3\beta_{2,010}\beta_{2,100})\gamma_{010} +(\beta_{1,010} \beta_{2,010} + \beta_{1,010}^2)\gamma_{100}\Big)
\nonumber\\
&&\hspace*{-5mm}\qquad
+\alpha_1^2\alpha_3 \Big(3\gamma_{100}^2 \gamma_{001} -3\beta_{1,001}\gamma_{100}^2 -3 (\beta_{3,100} +\beta_{1,100}) \gamma_{100}\gamma_{001} + (3\beta_{1,001}\beta_{1,100} +\beta_{1,001}
\nonumber\\
&&\hspace*{-5mm}\qquad
\times \beta_{3,100})\gamma_{100} +(\beta_{1,100}\beta_{3,100} + \beta_{3,100}^2)\gamma_{001} \Big)
+\alpha_1\alpha_3^2\Big(3\gamma_{001}^2 \gamma_{100} -3\beta_{3,100}\gamma_{001}^2  -3(\beta_{1,001}
\nonumber\\
&&\hspace*{-5mm}\qquad
+\beta_{3,001})\gamma_{001}\gamma_{100} + (\beta_{1,001}\beta_{3,100} +3\beta_{3,001}\beta_{3,100})\gamma_{001} +(\beta_{1,001} \beta_{3,001} + \beta_{1,001}^2)\gamma_{100}\Big)
\nonumber\\
&&\hspace*{-5mm}\qquad
+\alpha_2^2\alpha_3 \Big(3\gamma_{010}^2 \gamma_{001} -3\beta_{2,001}\gamma_{010}^2 -3 (\beta_{3,010} +\beta_{2,010}) \gamma_{010}\gamma_{001} + (3\beta_{2,001}\beta_{2,010} +\beta_{2,001}
\nonumber\\
&&\hspace*{-5mm}\qquad
\times \beta_{3,010})\gamma_{010} +(\beta_{2,010}\beta_{3,010} + \beta_{3,010}^2)\gamma_{001} \Big)
+\alpha_2\alpha_3^2\Big(3\gamma_{001}^2 \gamma_{010} -3\beta_{3,010}\gamma_{001}^2  -3(\beta_{2,001}
\nonumber\\
&&\hspace*{-5mm}\qquad
+\beta_{3,001})\gamma_{001}\gamma_{010} + (\beta_{2,001}\beta_{3,010} +3\beta_{3,001}\beta_{3,010})\gamma_{001} +(\beta_{2,001} \beta_{3,001} + \beta_{2,001}^2)\gamma_{010}\Big)
\nonumber\\
&&\hspace*{-5mm}\qquad
+\alpha_1\alpha_2\alpha_3\Big(6 \gamma_{001}\gamma_{010}\gamma_{100}
-3(\beta_{1,001} + \beta_{2,001})\gamma_{100}\gamma_{010}
-3(\beta_{1,010} + \beta_{3,010})\gamma_{100}\gamma_{001}
\nonumber\\
&&\hspace*{-5mm}\qquad
-3(\beta_{2,100} + \beta_{3,100})\gamma_{010}\gamma_{001}
+2\beta_{1,001}\beta_{1,010}\gamma_{100}
+2\beta_{2,001}\beta_{2,100}\gamma_{010}
+2\beta_{3,010}\beta_{3,100}\gamma_{001}
\vphantom{\Big(}\nonumber\\
&&\hspace*{-5mm}\qquad
+(\beta_{1,001}\beta_{3,010} +\beta_{1,010}\beta_{2,001})\gamma_{100}
+(\beta_{1,001}\beta_{2,100} +\beta_{2,001}\beta_{3,100})\gamma_{010}
+(\beta_{1,010}\beta_{3,100}
\vphantom{\Big(}\nonumber\\
&&\hspace*{-5mm}\qquad
+\beta_{2,100}\beta_{3,010})\gamma_{001}
\Big)
\Big]\nonumber\\
&&\hspace*{-5mm} + \Big(\frac{1}{2\varepsilon^2} + \frac{3}{2\varepsilon}\ln\frac{\Lambda}{\mu} + \ln^2\frac{\Lambda}{\mu}\Big)\Big[\alpha_1^3 \gamma_{100}\gamma_{200} + \alpha_2^3 \gamma_{010}\gamma_{020}
+ \alpha_3^3 \gamma_{001}\gamma_{002}
+\alpha_1^2\alpha_2\Big(\gamma_{010}\gamma_{200} + \gamma_{100}\nonumber\\
&&\hspace*{-5mm}\qquad
\times \gamma_{110}\Big) + \alpha_1\alpha_2^2\Big(\gamma_{020}\gamma_{100} +\gamma_{010} \gamma_{110} \Big) +\alpha_1^2\alpha_3\Big(\gamma_{001}\gamma_{200} + \gamma_{100}\gamma_{101}\Big) + \alpha_1\alpha_3^2\Big(\gamma_{002}\gamma_{100} \nonumber\\
&&\hspace*{-5mm}\qquad
+\gamma_{001} \gamma_{101} \Big)
+\alpha_2^2\alpha_3\Big(\gamma_{001}\gamma_{020} + \gamma_{010}\gamma_{011}\Big) + \alpha_2\alpha_3^2\Big(\gamma_{002}\gamma_{010} +\gamma_{001} \gamma_{011} \Big)
+\alpha_1 \alpha_2 \alpha_3 \Big(\gamma_{100}\nonumber\\
&&\hspace*{-5mm}\qquad \times \gamma_{011} + \gamma_{010} \gamma_{101} + \gamma_{001} \gamma_{110}\Big)
\Big]\nonumber\\
&&\hspace*{-5mm} - \Big(\frac{1}{6\varepsilon^2} + \frac{1}{2\varepsilon}\ln\frac{\Lambda}{\mu} +\frac{1}{2}\ln^2\frac{\Lambda}{\mu}\Big)\Big[2\alpha_1^3 \beta_{1,100}\gamma_{200}
+ 2\alpha_2^3 \beta_{2,010}\gamma_{020} + 2\alpha_3^3 \beta_{3,001}\gamma_{002}
\nonumber\\
&&\hspace*{-5mm}\qquad +\alpha_1^2\alpha_2\Big( (\beta_{1,100} +\beta_{2,100}) \gamma_{110} +2\beta_{1,010}\gamma_{200}\Big) + \alpha_1\alpha_2^2 \Big( (\beta_{1,010}+\beta_{2,010}) \gamma_{110}+2\beta_{2,100}\gamma_{020}\Big)
\nonumber\\
&&\hspace*{-5mm}\qquad
+\alpha_1^2\alpha_3\Big( (\beta_{1,100} +\beta_{3,100}) \gamma_{101} +2\beta_{1,001}\gamma_{200}\Big) + \alpha_1\alpha_3^2 \Big( (\beta_{1,001}+\beta_{3,001}) \gamma_{101}+2\beta_{3,100}\gamma_{002}\Big)
\nonumber\\
&&\hspace*{-5mm}\qquad
+\alpha_2^2\alpha_3\Big( (\beta_{2,010} +\beta_{3,010}) \gamma_{011} +2\beta_{2,001}\gamma_{020}\Big) + \alpha_2\alpha_3^2 \Big( (\beta_{2,001}+\beta_{3,001}) \gamma_{011}+2\beta_{3,010}\gamma_{002}\Big)
\nonumber\\
&&\hspace*{-5mm}\qquad
+ \alpha_1\alpha_2\alpha_3 \Big(\beta_{1,010}\gamma_{101} +\beta_{1,001} \gamma_{110} + \beta_{2,100}\gamma_{011} +\beta_{2,001}\gamma_{110} + \beta_{3,100}\gamma_{011} + \beta_{3,010}\gamma_{101}\Big)
\Big]\nonumber\\
&&\hspace*{-5mm} - \Big(\frac{1}{3\varepsilon^2} + \frac{1}{\varepsilon}\ln\frac{\Lambda}{\mu} + \frac{1}{2} \ln^2\frac{\Lambda}{\mu}\Big)\Big[\alpha_1^3 \beta_{1,200}\gamma_{100} + \alpha_2^3 \beta_{2,020}\gamma_{010}
+ \alpha_3^3 \beta_{3,002}\gamma_{001}
+\alpha_1^2\alpha_2\Big(  \beta_{1,110}\gamma_{100}\nonumber\\
&&\hspace*{-5mm}\qquad + \beta_{2,200}\gamma_{010} \Big) + \alpha_1\alpha_2^2\Big(\beta_{2,110}\gamma_{010} + \beta_{1,020}\gamma_{100}\Big)
+\alpha_1^2\alpha_3\Big(  \beta_{1,101}\gamma_{100} + \beta_{3,200}\gamma_{001} \Big) + \alpha_1\alpha_3^2\nonumber\\
&&\hspace*{-5mm}\qquad \times\Big(\beta_{3,101}\gamma_{001} + \beta_{1,002}\gamma_{100}\Big)
+\alpha_2^2\alpha_3\Big( \beta_{2,011}\gamma_{010} + \beta_{3,020}\gamma_{001} \Big) + \alpha_2\alpha_3^2\Big(\beta_{3,011}\gamma_{001} + \beta_{2,002}\nonumber\\
&&\hspace*{-5mm}\qquad \times \gamma_{010}\Big) +\alpha_1\alpha_2\alpha_3\Big(\beta_{1,011}\gamma_{100} +\beta_{2,101}\gamma_{010} + \beta_{3,110}\gamma_{001}\Big)
\Big]\nonumber\\
&&\hspace*{-5mm} - \Big(\frac{1}{3\varepsilon}+\ln\frac{\Lambda}{\mu}\Big)\Big[\alpha_1^3 \gamma_{300} +\alpha_2^3\gamma_{030} +\alpha_3^3\gamma_{003} +\alpha_1^2\alpha_2 \gamma_{210} +\alpha_1\alpha_2^2 \gamma_{120} +\alpha_1^2\alpha_3 \gamma_{201} +\alpha_1\alpha_3^2 \gamma_{102}\nonumber\\
&&\hspace*{-5mm}\qquad +\alpha_2^2\alpha_3 \gamma_{021} +\alpha_2\alpha_3^2 \gamma_{012} +\alpha_1\alpha_2\alpha_3\gamma_{111}\Big]
+ O(\alpha^4).
\end{eqnarray}

\section{Renormalization constants for theories with a single coupling in the MS-like renormalization schemes}
\hspace*{\parindent}\label{Appendix_Single_Coupling}

Let us demonstrate how the ordered exponential can be calculated for the theories which contain a single coupling $\alpha$ in the MS-like renormalization prescriptions. In this case the index $i$ takes only one value and, therefore, may be omitted.

It is convenient to choose the expression (\ref{Z_Result_Calculable}) as a starting point. Making the change of the integration variable $t\to t\alpha$ we obtain

\begin{equation}
\bigg(1- \int\limits^\wedge \frac{dt}{t\varepsilon}\Big( \frac{1}{t} \beta(t\alpha)\frac{\partial}{\partial\alpha} - \gamma(t\alpha)\Big)\bigg)^{-1}\cdot 1\,\Bigg|_{t=1} = \bigg(1- \int\limits^\wedge \frac{d\alpha}{\alpha \varepsilon}\Big( \beta(\alpha)\frac{\partial}{\partial\alpha} - \gamma(\alpha)\Big)\bigg)^{-1}\cdot 1.
\end{equation}

\noindent
To simplify this expression, we consider the product

\begin{eqnarray}\label{Auxiliary_Expression}
&& \bigg(1- \int\limits^\wedge \frac{d\alpha}{\alpha \varepsilon}\Big( \beta(\alpha)\frac{\partial}{\partial\alpha} - \gamma(\alpha)\Big)\bigg) \exp\bigg\{\int\limits_0^\alpha \frac{d\alpha\,\gamma(\alpha)}{\beta(\alpha)-\alpha\varepsilon}\bigg\}\nonumber\\
&&\qquad\qquad\qquad\qquad\qquad\qquad\qquad
= \bigg(1- \int\limits^\wedge \frac{d\alpha\,\gamma(\alpha)}{\beta(\alpha) - \alpha \varepsilon} \bigg) \exp\bigg\{\int\limits_0^\alpha \frac{d\alpha\,\gamma(\alpha)}{\beta(\alpha)-\alpha\varepsilon}\bigg\}. \qquad
\end{eqnarray}

\noindent
Introducing the function

\begin{equation}
f(\alpha)\equiv \int\limits_0^\alpha \frac{d\alpha\,\gamma(\alpha)}{\beta(\alpha)-\alpha\varepsilon},
\end{equation}

\noindent
the expression (\ref{Auxiliary_Expression}) can be rewritten in the form

\begin{equation}
\exp\left(f(\alpha)\right) - \int\limits_0^\alpha df(\alpha) \exp\left(f(\alpha)\right) = \exp\left(f(\alpha)\right) - \exp\left(f(\alpha)\right)\Big|_0^\alpha = \exp\left(f(0)\right) = 1.
\end{equation}

\noindent
This implies that we obtain the identity

\begin{equation}\label{Operator_Identity}
\bigg(1- \int\limits^\wedge \frac{dt}{t\varepsilon}\Big( \frac{1}{t} \beta(t\alpha)\frac{\partial}{\partial\alpha} - \gamma(t\alpha)\Big)\bigg)^{-1}\cdot 1\,\Bigg|_{t=1}
= \exp\bigg\{\int\limits_0^\alpha \frac{d\alpha\,\gamma(\alpha)}{\beta(\alpha)-\alpha\varepsilon}\bigg\}.
\end{equation}

\noindent
Substituting this result into Eq. (\ref{Z_Result_Calculable}) (for the particular case of a theory with a single coupling) we see that for the MS-like renormalization schemes the (integer-dimensional) renormalization constants can be presented in the form

\begin{equation}
Z\Big(\alpha,\varepsilon^{-1},\ln\frac{\Lambda}{\mu}\Big) = \exp\bigg\{\ln\frac{\Lambda}{\mu}\Big(\beta(\alpha)\frac{\partial}{\partial\alpha} - \gamma(\alpha)\Big)\bigg\}
\exp\bigg\{\int\limits_0^\alpha \frac{d\alpha\,\gamma(\alpha)}{\beta(\alpha)-\alpha\varepsilon}\bigg\}.
\end{equation}

\section{The expression for $\ln Z$}
\hspace*{\parindent}\label{Appendix_Ln_Z}

In this appendix we present the expression for $\ln Z$ because it is linear in the anomalous dimension and, therefore, may be more convenient for making explicit calculations. Unfortunately, we did not manage to present the general result in a such form where $\varepsilon$-poles and logarithms enter in a similar way. (However, it is simply a sum of terms linear in the anomalous dimension in Eq. (\ref{Z_TExponent}).) The expression for the logarithm of a renormalization constant can be written as

\begin{eqnarray}\label{Ln_Z_Result}
&& \ln Z\Big(\alpha,\varepsilon^{-1},\ln\frac{\Lambda}{\mu}\Big) = \exp\Big\{\ln\frac{\Lambda}{\mu}\sum\limits_{i=1}^n \beta_i(\alpha)\frac{\partial}{\partial\alpha_i}\Big\} \ln Z(\alpha,\varepsilon^{-1})
\nonumber\\
&&\qquad\qquad\qquad\qquad\qquad\qquad\qquad\qquad\qquad
- \sum\limits_{k=1}^\infty \frac{1}{k!} \ln^k\frac{\Lambda}{\mu} \Big(\sum\limits_{i=1}^n \beta_i(\alpha)\frac{\partial}{\partial\alpha_i}\Big)^{k-1} \gamma(\alpha),\qquad
\end{eqnarray}

\noindent
where $\ln Z(\alpha,\varepsilon^{-1})$ is the logarithm of the renormalization constant for the standard dimensional regularization/reduction given by the expression

\begin{eqnarray}\label{Z_Ln_DR}
&& \ln Z(\alpha,\varepsilon^{-1}) = - \int \limits_0^1 \frac{dt_1}{t_1\,\varepsilon} \gamma(t_1\alpha) - \int \limits_0^1 \frac{dt_1}{t_1^2\,\varepsilon}
\sum\limits_{i_1=1}^n \beta_{i_1}(t_1\alpha)\frac{\partial}{\partial\alpha_{i_1}} \int \limits_0^{t_1} \frac{dt_2}{t_2\,\varepsilon} \gamma(t_2\alpha)
\nonumber\\
&& \qquad\qquad\quad
- \int \limits_0^1 \frac{dt_1}{t_1^2\,\varepsilon}
\sum\limits_{i_1=1}^n \beta_{i_1}(t_1\alpha)\frac{\partial}{\partial\alpha_{i_1}}
\int \limits_0^{t_1} \frac{dt_2}{t_2^2\,\varepsilon} \sum\limits_{i_2=1}^n \beta_{i_2}(t_1\alpha)\frac{\partial}{\partial\alpha_{i_2}}
\int \limits_0^{t_2} \frac{dt_3}{t_3\,\varepsilon} \gamma(t_3\alpha)+\ldots\qquad
\vphantom{\int\limits_0^t}
\end{eqnarray}

\noindent
This result can be obtained from the equation

\begin{equation}
\Big(\frac{1}{t^2\varepsilon}\sum\limits_{i=1}^n \beta_i(t\alpha)\frac{\partial}{\partial\alpha_i} - \frac{\partial}{\partial t}\Big) \ln Z(t\alpha,\varepsilon^{-1}) =\frac{1}{t\varepsilon}\gamma(t\alpha),
\end{equation}

\noindent
which follows from Eq. (\ref{RG_Equation_DR}). Its solution, written with the help of the operator $\int\limits^\wedge dt$ defined by Eq. (\ref{Integral_Operator_Definition}), has the form

\begin{eqnarray}
&& \ln Z(t\alpha,\varepsilon^{-1}) = - \sum\limits_{k=0}^\infty\Big(\int\limits^\wedge \frac{dt}{t^2\varepsilon} \sum\limits_{i=1}^n \beta_i(t\alpha)\frac{\partial}{\partial\alpha_i}\Big)^k \int\limits^\wedge \frac{dt}{t \varepsilon}\gamma(t\alpha)\nonumber\\
&&\qquad\qquad\qquad\qquad\qquad\qquad\qquad = -\Big(1-\int\limits^\wedge \frac{dt}{t^2\varepsilon} \sum\limits_{i=1}^n \beta_i(t\alpha)\frac{\partial}{\partial\alpha_i}\Big)^{-1} \int\limits^\wedge \frac{dt}{t \varepsilon}\gamma(t\alpha).\qquad
\end{eqnarray}

\noindent
Similarly, the other terms in Eq. (\ref{Ln_Z_Result}) are obtained from Eq. (\ref{Ln_Z_RG_Equation}) if one takes into account the boundary condition at $\mu=\Lambda$, namely, $\ln Z(\alpha,\varepsilon^{-1},0) = \ln Z(\alpha,\varepsilon^{-1})$.

For the regularizations of the cutoff type from Eq. (\ref{Ln_Z_Result}) we should omit $\varepsilon$-poles, so that $\ln Z(\alpha,\varepsilon^{-1})\to 0$ and the result can be written as

\begin{equation}
\ln Z\Big(\alpha,\ln\frac{\Lambda}{\mu}\Big) =
- \sum\limits_{k=1}^\infty \frac{1}{k!} \ln^k\frac{\Lambda}{\mu} \Big(\sum\limits_{i=1}^n \beta_i(\alpha)\frac{\partial}{\partial\alpha_i}\Big)^{k-1} \gamma(\alpha).
\end{equation}

\noindent
Introducing the auxiliary variable $t$ and redefining the index of summation $k\to k+1$, this expression can be presented in a more beautiful form,

\begin{eqnarray}
&& \ln Z\Big(\alpha,\ln\frac{\Lambda}{\mu}\Big) =
- \sum\limits_{k=0}^\infty \frac{1}{k!} \int\limits_0^{\ln\Lambda/\mu} dt\, t^{k} \Big(\sum\limits_{i=1}^n \beta_i(\alpha)\frac{\partial}{\partial\alpha_i}\Big)^{k} \gamma(\alpha)
\nonumber\\
&& \qquad\qquad\qquad\qquad\qquad\qquad\qquad\qquad\quad
= - \int\limits_0^{\ln\Lambda/\mu} dt\,\exp\Big\{t \sum\limits_{i=1}^n \beta_i(\alpha)\frac{\partial}{\partial\alpha_i}\Big\}\,\gamma(\alpha).\qquad
\end{eqnarray}

\end{document}